\documentclass[%
reprint,
superscriptaddress,
amsmath,amssymb,
aps,
prx,
]{revtex4-2}

\usepackage[
CJKbookmarks=false,
colorlinks,
breaklinks=true,
linkcolor=blue,
anchorcolor=blue,
citecolor=blue,
urlcolor=blue,
]{hyperref}
\usepackage{subfigure}

\usepackage{textcomp}
\usepackage{graphicx}
\usepackage{dcolumn}
\usepackage{bm}

\begin{document}

\preprint{APS/123-QED}

\title{Multi-Purpose Architecture for Fast Reset and Protective Readout of Superconducting Qubits}


\author{Jiayu Ding}
\thanks{These authors contributed equally to this work.}
\affiliation{National Laboratory of Solid State Microstructures, School of Physics, Nanjing University, Nanjing 210093, China}
\affiliation{Beijing Academy of Quantum Information Sciences, Beijing 100193, China}
\affiliation{Synergetic Innovation Center of Quantum Information and Quantum Physics, University of Science and Technology of China, Hefei, Anhui 230026, China}

\author{Yulong Li}
\thanks{These authors contributed equally to this work.}
\affiliation{Beijing Academy of Quantum Information Sciences, Beijing 100193, China}

\author{He Wang}
\email{wanghe@baqis.ac.cn}
\affiliation{Beijing Academy of Quantum Information Sciences, Beijing 100193, China}

\author{Guangming Xue}
\email{xuegm@baqis.ac.cn}
\affiliation{Beijing Academy of Quantum Information Sciences, Beijing 100193, China}

\author{Tang Su}
\affiliation{Beijing Academy of Quantum Information Sciences, Beijing 100193, China}

\author{Chenlu Wang}
\affiliation{Beijing Academy of Quantum Information Sciences, Beijing 100193, China}

\author{Weijie Sun}
\affiliation{Beijing Academy of Quantum Information Sciences, Beijing 100193, China}

\author{Feiyu Li}
\affiliation{Beijing Academy of Quantum Information Sciences, Beijing 100193, China}

\author{Yujia Zhang}
\affiliation{National Laboratory of Solid State Microstructures, School of Physics, Nanjing University, Nanjing 210093, China}
\affiliation{Beijing Academy of Quantum Information Sciences, Beijing 100193, China}
\affiliation{Synergetic Innovation Center of Quantum Information and Quantum Physics, University of Science and Technology of China, Hefei, Anhui 230026, China}

\author{Yang Gao}
\affiliation{Beijing Academy of Quantum Information Sciences, Beijing 100193, China}
\affiliation{Institute of Physics, Chinese Academy of Sciences, Beijing 100190, China}
\affiliation{University of Chinese Academy of Sciences, Beijing 101408, China}

\author{Jun Peng}
\affiliation{State Key Laboratory of Millimeter Waves, School of Information Science and Engineering, Southeast University, Nanjing, China}

\author{Zhi Hao Jiang}
\affiliation{State Key Laboratory of Millimeter Waves, School of Information Science and Engineering, Southeast University, Nanjing, China}

\author{Yang Yu}
\affiliation{National Laboratory of Solid State Microstructures, School of Physics, Nanjing University, Nanjing 210093, China}
\affiliation{Synergetic Innovation Center of Quantum Information and Quantum Physics, University of Science and Technology of China, Hefei, Anhui 230026, China}

\author{Haifeng Yu}
\affiliation{Beijing Academy of Quantum Information Sciences, Beijing 100193, China}

\author{Fei Yan}
\email{yanfei@baqis.ac.cn}
\affiliation{Beijing Academy of Quantum Information Sciences, Beijing 100193, China}

\date{\today}

\begin{abstract}
The ability to fast reset a qubit state is crucial for quantum information processing. However, to actively reset a qubit requires engineering a pathway to interact with a dissipative bath, which often comes at the cost of reduced qubit protection from the environment. Here, we present a multi-purpose architecture that enables fast reset and protection of superconducting qubits during control and readout. In our design, two on-chip diplexers are connected by two transmission lines. The high-pass branch provides a flat passband for convenient allocation of readout resonators above the qubit frequencies, which is preferred for reducing measurement-induced state transitions. In the low-pass branch, we leverage a standing-wave mode below the maximum qubit frequency for a rapid reset. The qubits are located in the common stopband to inhibit dissipation during coherent operations. We demonstrate resetting a transmon qubit from its first excited state to the ground state in 100 ns, achieving a residual population of 2.7\%. The reset time may be further shortened to 27 ns by exploiting the coherent population inversion effect. We further extend the technique to resetting the qubit from its second excited state. Our approach promises scalable implementation of fast reset and qubit protection during control and readout, adding to the toolbox of dissipation engineering.
\end{abstract}

\maketitle


\section{\label{sec:introduction}INTRODUCTION}

Initialization of a quantum state is an important criterion for the physical implementation of quantum computation \cite{divincenzo2000physical}. It can be realized through either passive waiting until reaching thermal equilibrium or active reset using a dissipative bath \cite{reed2010fast,Geerlings2013Demonstrating,egger2018pulsed,Magnard2018Fast,zhou2021rapid,Sunada2022Fast,Yoshioka2023Active,maurya2023demand,Wang2024Efficient}. In addition to initialization, the reset operation is useful in quantum error correction \cite{mcewen2021removing, Battistel2021Hardware, miao2023overcoming, Marques2023All, lacroix2023fast, yang2024coupler}, preparation of stable quantum-correlated many-body states \cite{mi2024stable}, qubit reuse compilation \cite{DeCross2023Qubit} and simulation of open quantum systems \cite{barreiro2011open,kapit2020entanglement}.

Implementing a fast reset with superconducting qubits is a challenge. One approach is through real-time feedback conditioned on the outcome of a projective measurement \cite{2012Initialization, salathe2018low}. However, the performance of this approach is susceptible to measurement errors and feedback loop latency. Alternatively, a lossy readout resonator can drain excitation away from the qubits without the need for measurements or feedback operations \cite{Geerlings2013Demonstrating,egger2018pulsed,Magnard2018Fast,zhou2021rapid,mcewen2021removing, Battistel2021Hardware, miao2023overcoming, Marques2023All, lacroix2023fast}. To speed up the dissipative process while maintaining qubit coherence, an additional resonator known as a Purcell filter is often used as a spectral mask of the environment \cite{reed2010fast,jeffrey2014Fast,Walter2017Rapid,Sunada2024Photon,yen2024interferometric}. However, using the same resonator for both readout and reset imposes serious constraints. It requires a lower resonator frequency than the maximum qubit frequency to be compatible with the prevailing transmon qubit design \cite{mcewen2021removing}. Unfortunately, the resonator-below-qubit configuration is more vulnerable to measurement-induced state transitions due to the hybridization among highly excited states \cite{Sank2016Measurement, shillito2022Dynamics, Khezri2023Measurement}. In the alternate configuration, the phenomenon becomes less probable. Reset techniques compatible with the resonator-above-qubit configuration include parametric flux modulation \cite{zhou2021rapid, lacroix2023fast}, all-microwave driving \cite{Magnard2018Fast,Marques2023All}, bath engineering \cite{murch2012Cavity}, intrinsic Purcell filter \cite{Sunada2022Fast}, quantum circuit refrigerator \cite{yoshioka2021fast, sevriuk2022initial, Yoshioka2023Active}, and other methods. However, some of these methods may involve extra hardware costs or control complexity.

\begin{figure*}[htbp]
    \centering
    \includegraphics[width=\linewidth]{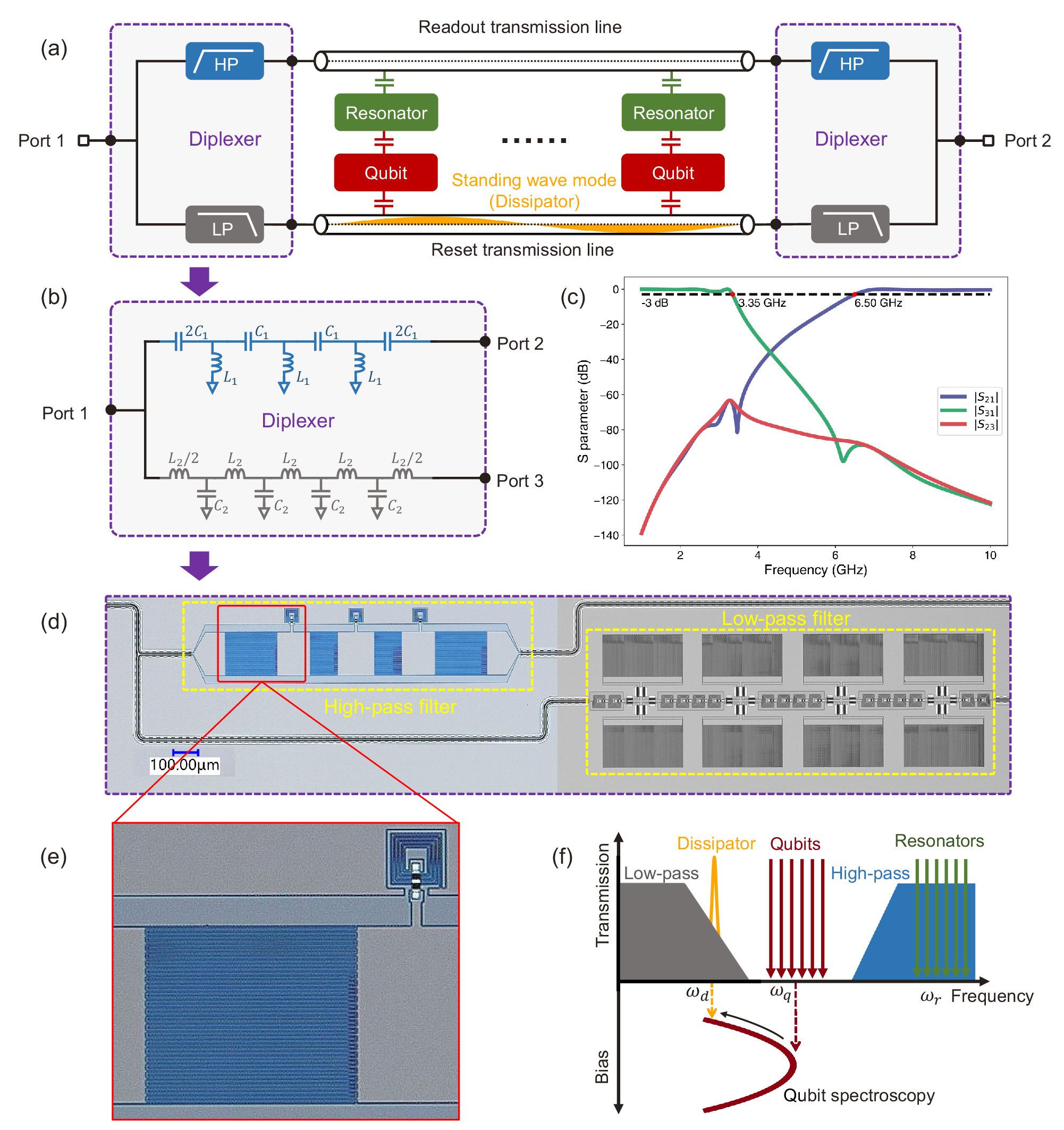}
    \subfigure{\label{fig:Layout:a}}
    \subfigure{\label{fig:Layout:b}}
    \subfigure{\label{fig:Layout:c}}
    \subfigure{\label{fig:Layout:d}}
    \subfigure{\label{fig:Layout:e}}
    \subfigure{\label{fig:Layout:f}}
    \caption{ Concept and design. (a) Schematic diagram of the design. Two on-chip diplexers, each composed of a high-pass filter and a low-pass filter, are connected through a pair of transmission lines. The high-pass channel couples to multiple readout resonators for transmitting readout signals, as in the conventional circuit QED architecture; the low-pass channel capacitively couples to multiple qubits (coupling capacitance of 6.3 fF) and is used as a dissipative channel. The dissipator is a standing-wave mode formed between the two low-pass filters. (b) Lumped circuit diagram of the diplexer. The design values for $C_1$, $L_1$, $C_2$, and $L_2$ are 0.266 pF, 0.660 nH, 1.809 pF, and 4.488 nH. (c) Scattering parameters of the diplexer. The port labels are with reference to (b). (d) Optical microscope image of the on-chip diplexer. The dashed boxes on the left and right indicate the high-pass and low-pass filters, respectively. (e) Enlarged view of the interdigitated capacitors and spiral inductors that make both filters. The widths of the fingers and gaps are all 2 \textmu m. The center of the spiral inductor is led out via an air bridge. (f) Frequency allocation of the low-pass and high-pass filters, the dissipator, the qubits, and the readout resonators. The qubits are protected from dissipation by the diplexers but can dissipate strongly through the dissipator mode.}
    \label{fig:Layout}
\end{figure*}

To address the aforementioned issues in one design, we introduce a multi-purpose architecture that employs a carefully engineered transmission line mode (the dissipator mode) that can couple to multiple superconducting qubits and serve as a common dissipative channel for fast resetting of the qubit states. The architecture features the use of on-chip diplexers that allow the physical separation of the readout and reset channels. This allows the readout resonator mode to be above the maximum qubit frequency and the dissipator mode to be below it, enabling both fast reset and fewer measurement-induced state transitions while preventing qubits from dissipating into the external circuitry during coherent operations. The experimental results show that, when the transmon qubit frequency is tuned in resonance with the dissipator mode, the population of the qubit in the first excited state decays with a characteristic rate of $\Gamma \approx 1/(21~\text{ns})$. With a reset duration of 100 ns (approximately $5 \Gamma^{-1}$), the residual error is approximately 2.7\%. Our design promises a hardware-efficient implementation of fast unconditional reset of multiple qubits while protecting the qubits from radiation loss and measurement-induced errors.

\section{\label{sec:principle}DEVICE DESIGN}

The key concept and circuit implementation of our design are illustrated in Fig.~\ref{fig:Layout}, which features the use of two symmetric on-chip diplexers, each composed of a low-pass filter with a cutoff frequency (-3 dB point) at 3.35 GHz and a high-pass filter with a cut-off frequency at 6.50 GHz [Fig.~\ref{fig:Layout:c}]. As shown in Fig.~\ref{fig:Layout:a}, connecting the two diplexers are two parallel transmission lines. The top high-pass line is coupled to multiple readout resonators for multiplexed readout. The bottom low-pass line (25 mm long) is coupled to multiple qubits. Each qubit is coupled to a dedicated control line
that carries both direct current and radio frequency signals \cite{pan2022engineering} (not shown in Fig.~\ref{fig:Layout}). A lossy standing-wave mode (4.23 GHz, full wavelength) is formed along the line, which we shall utilize for qubit reset. The diplexers are made up of lumped element capacitors and inductors [Fig.~\ref{fig:Layout:b}]. We use interdigitated capacitors and spiral inductors [Fig.~\ref{fig:Layout:e}] upon implementation, which produces a dimension of 3.7 $\times$ 0.8 $\mathrm{mm^2}$ in the final layout [Fig.~\ref{fig:Layout:d}]. 

The purpose of the design is multifold.
As illustrated in Fig.~\ref{fig:Layout:f}, the filters are designed such that the resonators ($\omega_\mathrm{r}/2\pi \sim$7.2--7.4 GHz fall within the passband of the high-pass filter, leading to a uniform external coupling. Meanwhile, the qubits ($\omega_\mathrm{q}/2\pi \sim 4.8$ GHz) are allocated within the common stopband of the high-pass and low-pass filters. This way, the qubits are protected from dissipation into the external circuitry as they see an unmatched impedance at the frequency of coherent operations. Unlike conventional bandpass or notch filters, both the qubits and resonators now have a large bandwidth for convenient frequency allocation, promising a greater readout multiplicity.

Over the low-pass line, a standing-wave mode can be formed between the two low-pass filters above their cutoff frequency where reflection dominates, reminiscent of a Fabry-Perot cavity formed by two reflective mirrors. With the finite transmission of the low-pass filter ($S_{21} \approx -30$\,dB) near the mode frequency at $\omega_\mathrm{d}/2\pi = 4.23$ GHz (design value), one can expect enhanced dissipation around this mode. Therefore, tuning the frequency of a tunable transmon qubit down to the dissipator frequency will deplete excitations in the qubit. In addition, the transmission between the low-pass and high-pass paths is suppressed by at least 60 dB over a wide frequency range (1–10 GHz), as shown in Fig.~\ref{fig:Layout:c}.

In the qubit-dissipator system, the effective decay rate $\Gamma$ of the qubit is modified by the lossy resonator --- known as the Purcell effect --- and can be given by \cite{heinsoo2018rapid,zhou2021rapid}
\begin{equation}\label{eq:purcell}
\Gamma = \frac{1}{2}\left(\kappa_{\mathrm{d}}-\operatorname{Re}\left\{\sqrt{-16 g^2+\left(\kappa_{\mathrm{d}}-2 i \Delta_{\mathrm{qd}}\right)^2}\right\}\right)
\end{equation}
where $\kappa_\mathrm{d}$ is the decay rate or linewidth of the dissipator, $g$ is the coupling strength between the qubit and the dissipator, and $\Delta_\mathrm{qd} = \omega_\mathrm{q} - \omega_\mathrm{d}$ is the detuning between the qubit and the dissipator. The qubit relaxation rate peaks when resonant with the dissipator ($\Delta_\mathrm{qd} = 0$). Note that the maximum decay rate $\kappa_{\mathrm{d}}/2$ can always be achieved at positions satisfying $g \geq \kappa_{\mathrm{d}}/4$. We target parameters of $\kappa_\mathrm{d}/2\pi \sim 15$ MHz and $g/2\pi \geq 4$ MHz to achieve a fast reset rate of $\Gamma \sim 21~\text{ns}^{-1}$. The linewidth of the readout resonator is designed to be $\kappa_{\mathrm{r}}/2\pi \sim 3.0$ MHz. The filters are designed to suppress the spontaneous emission of qubits through both the low-pass and high-pass paths. The qubit decay rate would be suppressed by a matched Purcell filter by a factor of $(1-|S_{11}|^2)/|1-S_{11}|^2$, where $S_{11}$ is the reflection parameter of the filter \cite{cleland2019mechanical}. Based on simulations of the circuit [Fig.~\ref{fig:Layout:b}], the Purcell factor is assumed to be below -40 dB for the low-pass filter when the qubit is above 4.2 GHz and to be below -30 dB for the high-pass filter when the qubit is below 5.0 GHz.

\begin{figure}[tb]
    \centering
    \includegraphics[width=\linewidth]{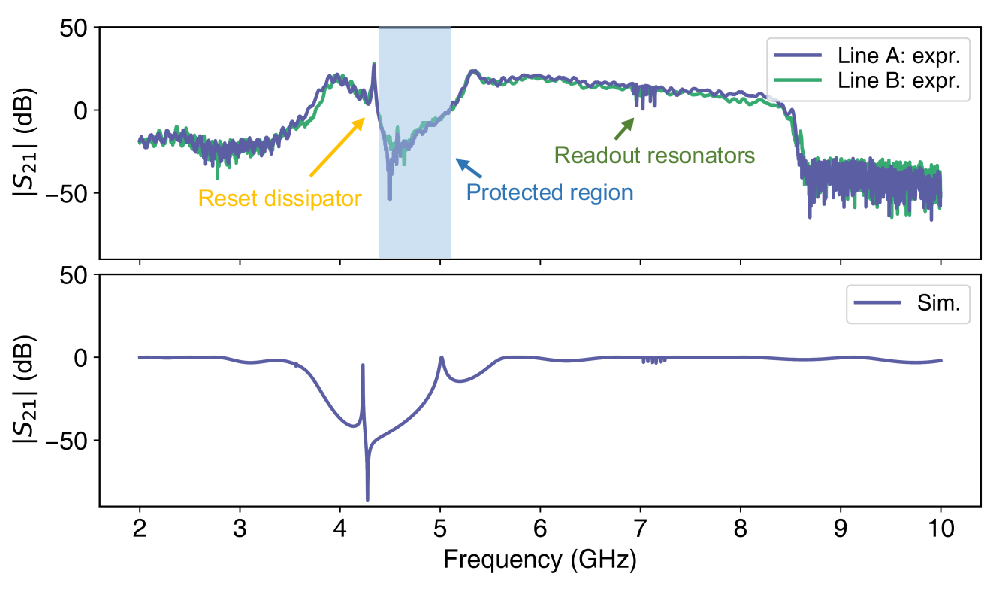}
    \subfigure{\label{fig:diplexer:a}}
    \subfigure{\label{fig:diplexer:b}}
    \caption{ The $S_{21}$ parameter. Experimentally measured and simulated transmission coefficients of the whole structure in Fig.~\ref{fig:Layout:a}. Two lines with identical designs are measured, showing good consistency. The envelope observed between 4 and 8 GHz is due to the bandwidth of the high electron mobility transistor amplifier used in the measurement chain. Note that the experimental results also reflect all the filtering, attenuation and amplification in the measurement chain.}
    \label{fig:s21}
\end{figure}

\section{\label{sec:experiment}EXPERIMENTAL RESULTS}

\begin{figure*}[tb]
  \centering
  \includegraphics[width=\linewidth]{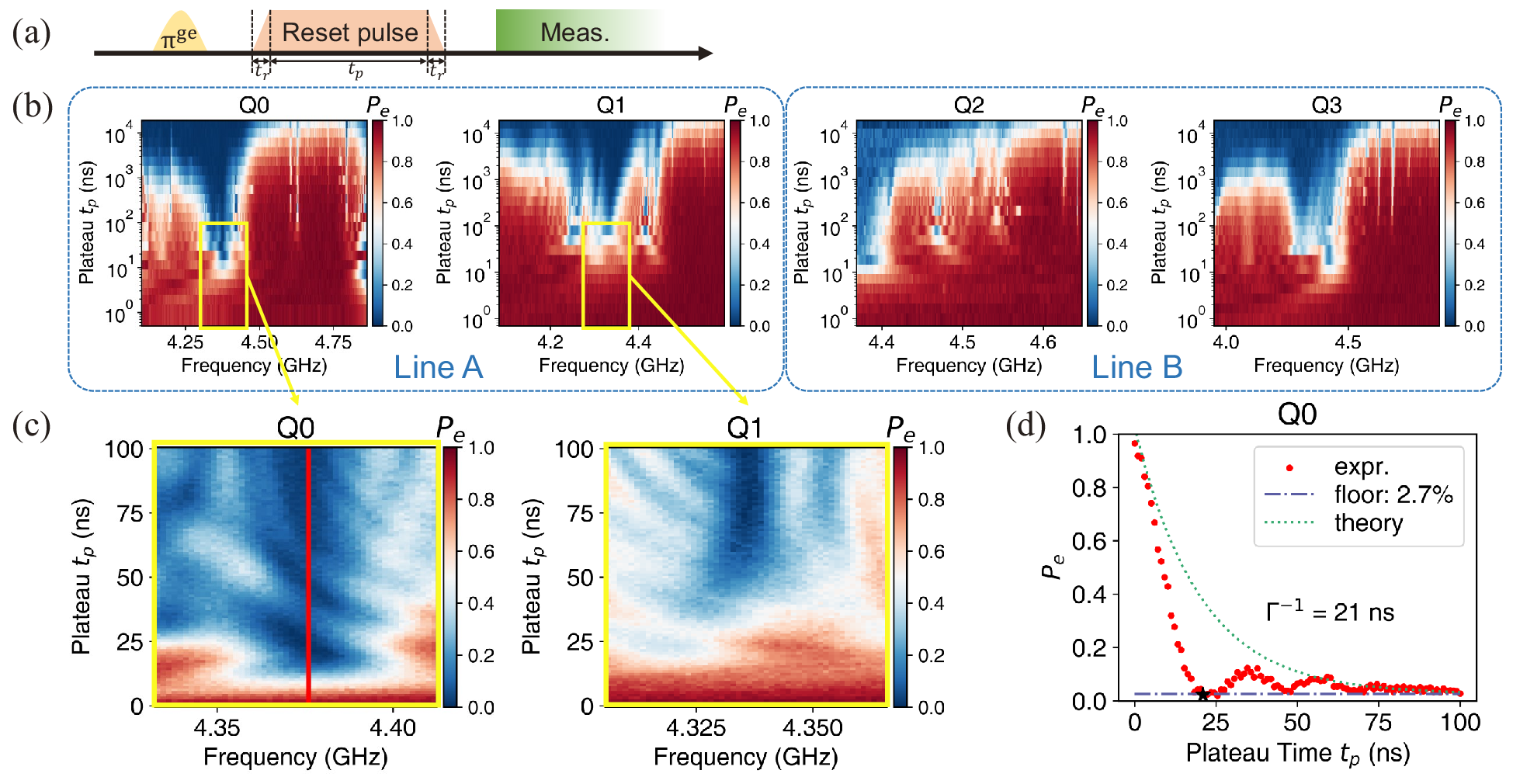}
  \subfigure{\label{fig:ResetResult:a}}
  \subfigure{\label{fig:ResetResult:b}}
  \subfigure{\label{fig:ResetResult:c}}
  \subfigure{\label{fig:ResetResult:d}}
  \caption{ Calibrating the $|e\rangle\mbox{-}|g\rangle$ reset operation. Calibrating the reset pulse. (a) The pulse sequence for calibrating a reset operation. The qubits are prepared in their first excited states and then frequency-tuned (rise time $t_r=2$\,ns) to be in resonance with the dissipator mode for a variable time ($t_\mathrm{p}$) before measurement. (b) The plots show the measured excited population for a variable $t_\mathrm{p}$ and qubit frequency which is adjusted by the pulse amplitude. Q0 and Q1 are coupled to a shared dissipator, while Q2 and Q3 are coupled to the other one. The strong dissipator is located around 4.3-4.4\,GHz. (c) Simultaneous reset of two qubits using a single reset line. The detailed scan is from the yellow boxed region in (b). Q0 with fringing patterns due to a strong qubit-dissipator coupling. (d) A linecut in (c) as indicated by the red solid line. The population reaches its first minimum at $t_\mathrm{p} = 23$ ns.}
  \label{fig:ResetResult}
\end{figure*}

\begin{figure}[t]
    \centering
    \includegraphics[width=1\linewidth]{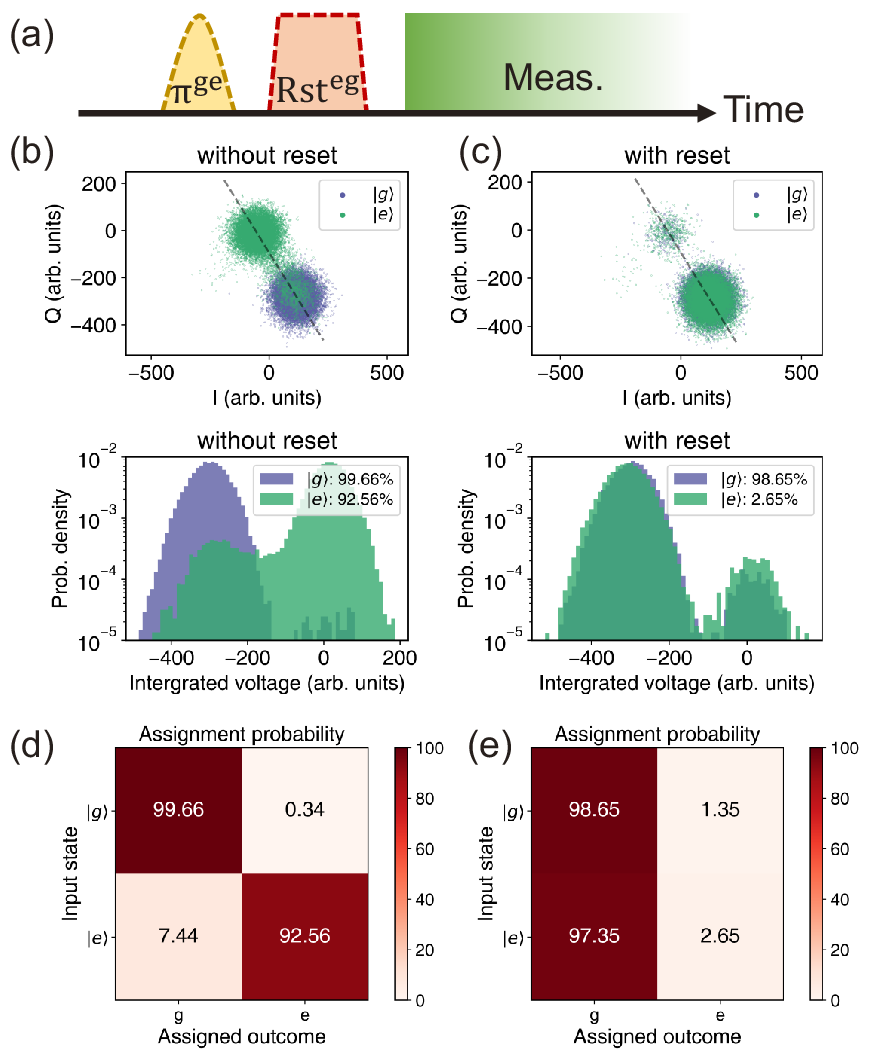}
    \subfigure{\label{fig:SingleShot10:a}}
    \subfigure{\label{fig:SingleShot10:b}}
    \subfigure{\label{fig:SingleShot10:c}}
    \subfigure{\label{fig:SingleShot10:d}}
    \subfigure{\label{fig:SingleShot10:e}}
    \caption{ Efficiency of the $|e\rangle\mbox{-}|g\rangle$ reset.  (a) Time sequence of the experiment. We use a plateau time $t_\mathrm{p}=200$ ns for the reset pulse. (b-c) Comparison of single-shot results for different initial states with and without the reset. (d-e) Comparison of the assignment probability in both cases. The reset gate reduces the population of the $|e\rangle$ state from 92.56\% to 2.65\%. }
    \label{fig:SingleShot10}
\end{figure}

\begin{figure}[t]
    \centering
    \includegraphics[width=1\linewidth]{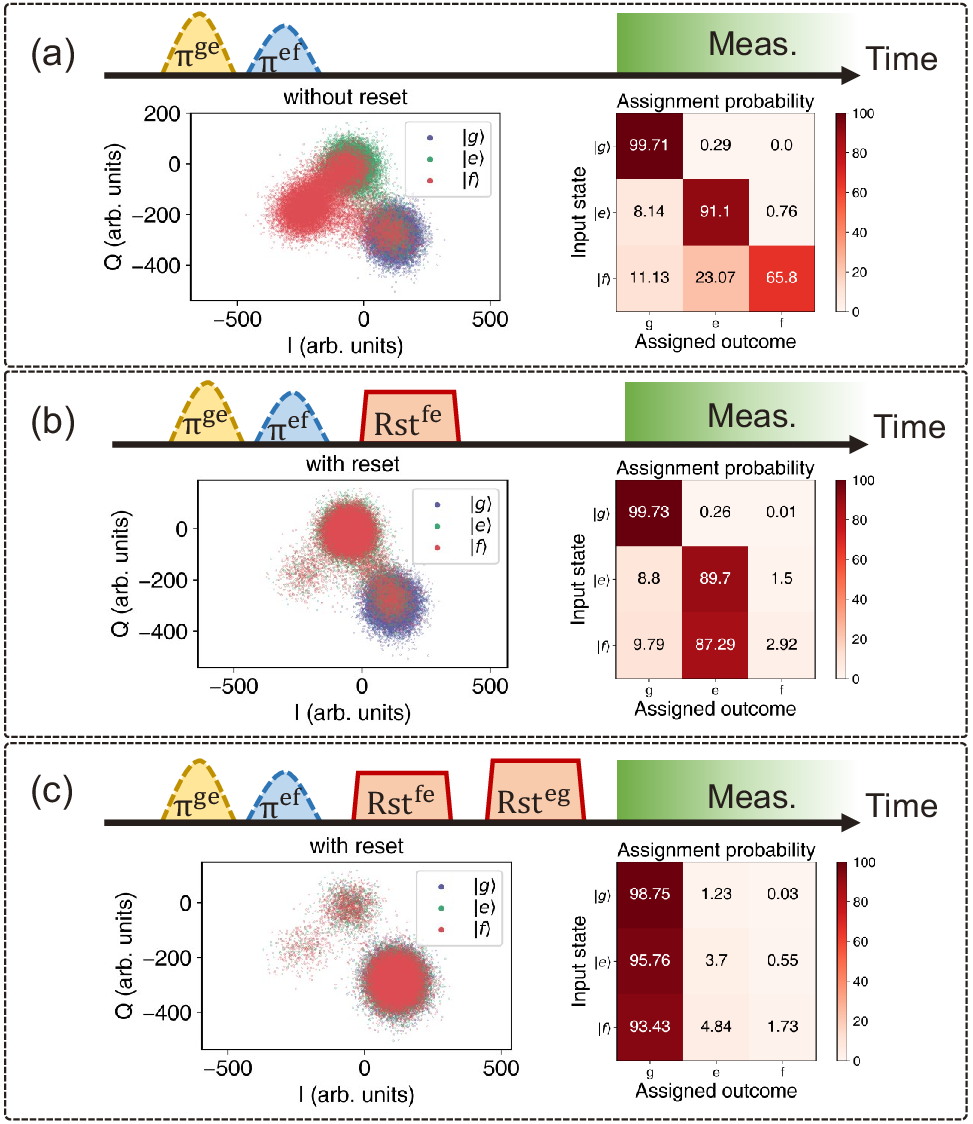}
    \subfigure{\label{fig:SingleShot21:a}}
    \subfigure{\label{fig:SingleShot21:b}}
    \subfigure{\label{fig:SingleShot21:c}}
    \caption{ Efficiency of concatenated reset gates. The three-state single-shot measurement results and the corresponding assignment probability matrix (a) without any reset gate, (b) with the $|f\rangle\mbox{-}|e\rangle$ reset gate, and (c) with concatenated $|f\rangle\mbox{-}|e\rangle$ and $|e\rangle\mbox{-}|g\rangle$ reset gates. }
    \label{fig:SingleShot21}
\end{figure}

In Fig.~\ref{fig:s21}, we plot the $S_{21}$ signal between the input and output ports of the full measurement chain using a network analyzer and compare them with finite-element simulations of the transmission between the two ports as depicted in Fig.~\ref{fig:Layout:a}. Two identical lines are measured in our experiment. In both plots, the dissipator, readout resonators, and protected region are clearly identifiable. The discrepancy in the filter frequencies is mainly due to imperfect modeling and fabrication variations. The linewidth of the dissipator extracted from the S parameters is $\kappa_\mathrm{d}/2\pi \sim$ 10--15 MHz.

We then calibrate the reset dynamics using the time sequence described in Fig.~\ref{fig:ResetResult:a}. The qubits are first prepared in their first excited states $|e\rangle$, and then frequency-tuned using a square flux pulse before projective measurement. By scanning the pulse amplitude and duration, it can be seen that the qubits dissipate about 3 orders of magnitude faster when biased near the dissipator at 4.3-4.4 GHz [Fig.~\ref{fig:ResetResult:b}], which is about 200 MHz above the design value due to imperfect modeling and fabrication variations. Note that there is uncertainty in the nominal frequency as the qubit spectra are not visible near the dissipator. In our design, each reset line couples to six transmon qubits. We carefully calibrate two qubits from each line (Q0 and Q1 from Line $A$, and Q2 and Q3 from Line $B$) with results shown in Fig.~\ref{fig:ResetResult:b}. The measured energy relaxation time $T_1$ of the qubits at the maximum frequency is 30--40 \textmu s. Then, we perform the simultaneous reset of two qubits, both reaching equilibrium after approximately 100 ns [Fig.~\ref{fig:ResetResult:c}].
For Q0, a finer scan in the vicinity of the dissipator resolves a fringing pattern [Fig.~\ref{fig:ResetResult:d}]. This is because the excitation can swap between the qubit and the dissipator before being finally dissipated given a relatively strong coupling $g/2\pi \approx 10$ MHz by design. As shown in Fig.~\ref{fig:ResetResult:d}, the qubit excitation oscillates while rapidly dropping to a baseline (2.7\%) within approximately 100 ns. The envelope decay is in good agreement with the theoretical value $\Gamma^{-1} = 21\text{ ns}$ according to Eq.~(\ref{eq:purcell}). On the other hand, the reset gate can be further shortened by exploiting the excitation swapping effect, for example, to 27 ns at the first minimum \cite{zhou2021rapid}. 

Next, we benchmark the efficiency of the calibrated $|e\rangle\mbox{-}|g\rangle$ reset gate of Q0 by initializing the qubit in different states [Fig.~\ref{fig:SingleShot10:a}].
The single-shot distribution plots [Figs.~\ref{fig:SingleShot10:b}-\ref{fig:SingleShot10:c}] and assignment matrices [Figs.~\ref{fig:SingleShot10:d}-\ref{fig:SingleShot10:e}] show that the majority of the excited-state population is depleted. We use a pulse plateau of $t_\mathrm{p}=200$ ns (approximately $10\Gamma^{-1}$) for the reset pulse to ensure a steady state. The reset gate reduces the excited-state population from 92.56\% to 2.65\%, consistent with the experiment shown in Fig.~\ref{fig:ResetResult:d}. The residual error may be due to thermal effects, leakage from state preparation or frequency tuning of the qubit, or readout error. However, a detailed and accurate analysis to identify the budget of residual errors is still being explored.

At the maximum qubit frequency of 4.86 GHz (far detuned from the dissipator), the excited-state population is approximately 0.34\% [Fig.~\ref{fig:SingleShot10:d}], corresponding to an effective noise temperature of $T_{\text {bath }}=41$ mK, according to the Boltzmann distribution $p_e / p_g=\exp \left(-\hbar \omega_q / k_{\mathrm{B}} T_{\text{bath}}\right)$, where $\hbar$ is the reduced Planck constant and $k_{\mathrm{B}}$ is the Boltzmann constant. However, when the qubit, initially in the ground state, is tuned to resonate with the dissipator at 4.37 GHz and held for sufficient time (approximately $10 \Gamma^{-1}$), the excited-state population reaches about 1.35\% [Fig.~\ref{fig:SingleShot10:e}], corresponding to an effective noise temperature of 49 mK for the dissipator. Although the dilution refrigerator provides a temperature below 20 mK, measured device temperature is typically higher due to residual radiation inside the refrigerator \cite{corcoles2011protecting}. The higher equilibrium temperature of the dissipator may be caused by additional noise from higher-temperature plates via the feedlines. This can be improved by using standalone termination for the dissipator line on the chip or the package.

We further test the efficiency of the reset gate applied to the qubit in its second excited state. Similar to the calibration of the $|e\rangle\mbox{-}|g\rangle$ reset gate, we apply a flux pulse that adjusts the qubit frequency to let the $|e\rangle\mbox{-}|f\rangle$ transition be resonant with the dissipator to depopulate the second excited state. To benchmark the $|f\rangle\mbox{-}|e\rangle$ reset gate, we first perform three-state discrimination and show the assignment probability matrix in [Fig.~\ref{fig:SingleShot21:a}]. The population of the $|f\rangle$ state rises to 0.76\% during the preparation of the $|e\rangle$ state [Fig.~\ref{fig:SingleShot21:a}], indicating leakage due to an unoptimized state preparation process. Then, we study the performance of a single $|f\rangle\mbox{-}|e\rangle$ reset operation [Fig.~\ref{fig:SingleShot21:b}]. The population of an initial $|f\rangle$ state decreases from 65.8\% to 2.92\%, while the population of an initial $|e\rangle$ state increases from 0.76\% to 1.5\%, as a consequence of the same thermal effect as in the $|e\rangle\mbox{-}|g\rangle$ reset gate. We note that the $|f\rangle\mbox{-}|e\rangle$ reset gate can be used for leakage reduction during quantum error correction, provided that the residual error can be further suppressed. We finally show the performance of concatenated $|f\rangle\mbox{-}|e\rangle$ and $|e\rangle\mbox{-}|g\rangle$ reset gates [Fig.~\ref{fig:SingleShot21:c}]. The cascade of the two reset gates reduces the population of the $|f\rangle$ state from 65.8\% to 1.73\%, showcasing the dissipator's capability to reset qubits from even higher excited states using the same mechanism. Since the transmon qubit has a negative anharmonicity, it is possible to go through a ladder of biases and reset the qubit step-by-step from a high-energy state to the ground state. The reset pulse can also be integrated into one pulse in piecewise steps. 

\section{\label{sec:conclusion}CONCLUSIONS}
To summarize, we have introduced an architecture based on superconducting quantum circuits. The architecture can satisfy multiple functionalities simultaneously. By separating qubit reset and readout channels using on-chip diplexers, the architecture provides substantial flexibility in circuit design, enabling protection for qubits from dissipating into the lossy circuitry, convenient allocation of qubit and resonator frequencies, a flat passband for the resonators, a qubit-below-resonator configuration for less measurement-induced state transitions, and, most importantly, fast reset under these constraints. The residual error can be improved by lowering the effective device temperature, minimizing control errors, and elevating the frequencies of the dissipator. Note that the dissipator channel may be terminated with a matched impedance separately without diplexing, preventing additional thermal noise from higher-temperature stages via cables. In the future, the filter design can be optimized for compactness, a narrow transition band, and precise alignment of the dissipator frequency near the minimum qubit frequency. On the experimental side, the performance of resetting more qubits simultaneously or serially is left to test. The scheme is promising for scalable implementation in quantum error correction, enabling the resetting of syndrome qubits and reducing leakage in data qubits. 
Furthermore, the engineered dissipative bath is a valuable resource and could be integrated as an integral part of dissipation engineering, which attracts significant attention in quantum information science \cite{harrington2022engineered}, such as the preparation of dissipative stabilized quantum states \cite{shankar2013autonomously,mi2024stable}, the realization of autonomous quantum error correction \cite{kerckhoff2010designing,kapit2016hardware,reiter2017dissipative,gertler2021protecting,Marquet2024Autoparametric} and the study of quantum transport \cite{harris2017quantum,mi2024stable}. Given the flexibility of the architecture, we believe our versatile design will find broader practical applications in superconducting quantum information processing.

\begin{acknowledgments}
The authors would like to thank Ji Chu and Yasunobu Nakamura for their valuable comments. This work was supported by the National Natural Science Foundation of China (Grants No. 92365206, No. 12322413, and No. 92476206), Innovation Program for Quantum Science and Technology (Grants No.2021ZD0301802 and No. 2021ZD0301702), and NSF of Jiangsu Province (Grants No. BE2021015-1 and No. BK20232002).
\end{acknowledgments}



\bibliography{apssamp}

\providecommand{\noopsort}[1]{}\providecommand{\singleletter}[1]{#1}%
\begin{thebibliography}{44}%
\makeatletter
\providecommand \@ifxundefined [1]{%
 \@ifx{#1\undefined}
}%
\providecommand \@ifnum [1]{%
 \ifnum #1\expandafter \@firstoftwo
 \else \expandafter \@secondoftwo
 \fi
}%
\providecommand \@ifx [1]{%
 \ifx #1\expandafter \@firstoftwo
 \else \expandafter \@secondoftwo
 \fi
}%
\providecommand \natexlab [1]{#1}%
\providecommand \enquote  [1]{``#1''}%
\providecommand \bibnamefont  [1]{#1}%
\providecommand \bibfnamefont [1]{#1}%
\providecommand \citenamefont [1]{#1}%
\providecommand \href@noop [0]{\@secondoftwo}%
\providecommand \href [0]{\begingroup \@sanitize@url \@href}%
\providecommand \@href[1]{\@@startlink{#1}\@@href}%
\providecommand \@@href[1]{\endgroup#1\@@endlink}%
\providecommand \@sanitize@url [0]{\catcode `\\12\catcode `\$12\catcode `\&12\catcode `\#12\catcode `\^12\catcode `\_12\catcode `\%12\relax}%
\providecommand \@@startlink[1]{}%
\providecommand \@@endlink[0]{}%
\providecommand \url  [0]{\begingroup\@sanitize@url \@url }%
\providecommand \@url [1]{\endgroup\@href {#1}{\urlprefix }}%
\providecommand \urlprefix  [0]{URL }%
\providecommand \Eprint [0]{\href }%
\providecommand \doibase [0]{https://doi.org/}%
\providecommand \selectlanguage [0]{\@gobble}%
\providecommand \bibinfo  [0]{\@secondoftwo}%
\providecommand \bibfield  [0]{\@secondoftwo}%
\providecommand \translation [1]{[#1]}%
\providecommand \BibitemOpen [0]{}%
\providecommand \bibitemStop [0]{}%
\providecommand \bibitemNoStop [0]{.\EOS\space}%
\providecommand \EOS [0]{\spacefactor3000\relax}%
\providecommand \BibitemShut  [1]{\csname bibitem#1\endcsname}%
\let\auto@bib@innerbib\@empty
\bibitem [{\citenamefont {DiVincenzo}(2000)}]{divincenzo2000physical}%
  \BibitemOpen
  \bibfield  {author} {\bibinfo {author} {\bibfnamefont {D.~P.}\ \bibnamefont {DiVincenzo}},\ }\bibfield  {title} {\bibinfo {title} {The physical implementation of quantum computation},\ }\href {https://doi.org/10.1002/1521-3978(200009)48:9/11<771::AID-PROP771>3.0.CO;2-E} {\bibfield  {journal} {\bibinfo  {journal} {Fortschr. Phys.}\ }\textbf {\bibinfo {volume} {48}},\ \bibinfo {pages} {771} (\bibinfo {year} {2000})}\BibitemShut {NoStop}%
\bibitem [{\citenamefont {Reed}\ \emph {et~al.}(2010)\citenamefont {Reed}, \citenamefont {Johnson}, \citenamefont {Houck}, \citenamefont {DiCarlo}, \citenamefont {Chow}, \citenamefont {Schuster}, \citenamefont {Frunzio},\ and\ \citenamefont {Schoelkopf}}]{reed2010fast}%
  \BibitemOpen
  \bibfield  {author} {\bibinfo {author} {\bibfnamefont {M.~D.}\ \bibnamefont {Reed}}, \bibinfo {author} {\bibfnamefont {B.~R.}\ \bibnamefont {Johnson}}, \bibinfo {author} {\bibfnamefont {A.~A.}\ \bibnamefont {Houck}}, \bibinfo {author} {\bibfnamefont {L.}~\bibnamefont {DiCarlo}}, \bibinfo {author} {\bibfnamefont {J.~M.}\ \bibnamefont {Chow}}, \bibinfo {author} {\bibfnamefont {D.~I.}\ \bibnamefont {Schuster}}, \bibinfo {author} {\bibfnamefont {L.}~\bibnamefont {Frunzio}},\ and\ \bibinfo {author} {\bibfnamefont {R.~J.}\ \bibnamefont {Schoelkopf}},\ }\bibfield  {title} {\bibinfo {title} {Fast reset and suppressing spontaneous emission of a superconducting qubit},\ }\href {https://doi.org/10.1063/1.3435463} {\bibfield  {journal} {\bibinfo  {journal} {Appl. Phys. Lett.}\ }\textbf {\bibinfo {volume} {96}} (\bibinfo {year} {2010})}\BibitemShut {NoStop}%
\bibitem [{\citenamefont {Geerlings}\ \emph {et~al.}(2013)\citenamefont {Geerlings}, \citenamefont {Leghtas}, \citenamefont {Pop}, \citenamefont {Shankar}, \citenamefont {Frunzio}, \citenamefont {Schoelkopf}, \citenamefont {Mirrahimi},\ and\ \citenamefont {Devoret}}]{Geerlings2013Demonstrating}%
  \BibitemOpen
  \bibfield  {author} {\bibinfo {author} {\bibfnamefont {K.}~\bibnamefont {Geerlings}}, \bibinfo {author} {\bibfnamefont {Z.}~\bibnamefont {Leghtas}}, \bibinfo {author} {\bibfnamefont {I.~M.}\ \bibnamefont {Pop}}, \bibinfo {author} {\bibfnamefont {S.}~\bibnamefont {Shankar}}, \bibinfo {author} {\bibfnamefont {L.}~\bibnamefont {Frunzio}}, \bibinfo {author} {\bibfnamefont {R.~J.}\ \bibnamefont {Schoelkopf}}, \bibinfo {author} {\bibfnamefont {M.}~\bibnamefont {Mirrahimi}},\ and\ \bibinfo {author} {\bibfnamefont {M.~H.}\ \bibnamefont {Devoret}},\ }\bibfield  {title} {\bibinfo {title} {Demonstrating a driven reset protocol for a superconducting qubit},\ }\href {https://doi.org/10.1103/PhysRevLett.110.120501} {\bibfield  {journal} {\bibinfo  {journal} {Phys. Rev. Lett.}\ }\textbf {\bibinfo {volume} {110}},\ \bibinfo {pages} {120501} (\bibinfo {year} {2013})}\BibitemShut {NoStop}%
\bibitem [{\citenamefont {Egger}\ \emph {et~al.}(2018)\citenamefont {Egger}, \citenamefont {Werninghaus}, \citenamefont {Ganzhorn}, \citenamefont {Salis}, \citenamefont {Fuhrer}, \citenamefont {M{\"u}ller},\ and\ \citenamefont {Filipp}}]{egger2018pulsed}%
  \BibitemOpen
  \bibfield  {author} {\bibinfo {author} {\bibfnamefont {D.~J.}\ \bibnamefont {Egger}}, \bibinfo {author} {\bibfnamefont {M.}~\bibnamefont {Werninghaus}}, \bibinfo {author} {\bibfnamefont {M.}~\bibnamefont {Ganzhorn}}, \bibinfo {author} {\bibfnamefont {G.}~\bibnamefont {Salis}}, \bibinfo {author} {\bibfnamefont {A.}~\bibnamefont {Fuhrer}}, \bibinfo {author} {\bibfnamefont {P.}~\bibnamefont {M{\"u}ller}},\ and\ \bibinfo {author} {\bibfnamefont {S.}~\bibnamefont {Filipp}},\ }\bibfield  {title} {\bibinfo {title} {Pulsed reset protocol for fixed-frequency superconducting qubits},\ }\href {https://doi.org/https://doi.org/10.1103/PhysRevApplied.10.044030} {\bibfield  {journal} {\bibinfo  {journal} {Phys. Rev. Appl.}\ }\textbf {\bibinfo {volume} {10}},\ \bibinfo {pages} {044030} (\bibinfo {year} {2018})}\BibitemShut {NoStop}%
\bibitem [{\citenamefont {Magnard}\ \emph {et~al.}(2018)\citenamefont {Magnard}, \citenamefont {Kurpiers}, \citenamefont {Royer}, \citenamefont {Walter}, \citenamefont {Besse}, \citenamefont {Gasparinetti}, \citenamefont {Pechal}, \citenamefont {Heinsoo}, \citenamefont {Storz}, \citenamefont {Blais},\ and\ \citenamefont {Wallraff}}]{Magnard2018Fast}%
  \BibitemOpen
  \bibfield  {author} {\bibinfo {author} {\bibfnamefont {P.}~\bibnamefont {Magnard}}, \bibinfo {author} {\bibfnamefont {P.}~\bibnamefont {Kurpiers}}, \bibinfo {author} {\bibfnamefont {B.}~\bibnamefont {Royer}}, \bibinfo {author} {\bibfnamefont {T.}~\bibnamefont {Walter}}, \bibinfo {author} {\bibfnamefont {J.-C.}\ \bibnamefont {Besse}}, \bibinfo {author} {\bibfnamefont {S.}~\bibnamefont {Gasparinetti}}, \bibinfo {author} {\bibfnamefont {M.}~\bibnamefont {Pechal}}, \bibinfo {author} {\bibfnamefont {J.}~\bibnamefont {Heinsoo}}, \bibinfo {author} {\bibfnamefont {S.}~\bibnamefont {Storz}}, \bibinfo {author} {\bibfnamefont {A.}~\bibnamefont {Blais}},\ and\ \bibinfo {author} {\bibfnamefont {A.}~\bibnamefont {Wallraff}},\ }\bibfield  {title} {\bibinfo {title} {Fast and unconditional all-microwave reset of a superconducting qubit},\ }\href {https://doi.org/10.1103/PhysRevLett.121.060502} {\bibfield  {journal} {\bibinfo  {journal} {Phys. Rev. Lett.}\ }\textbf {\bibinfo {volume} {121}},\ \bibinfo {pages} {060502}
  (\bibinfo {year} {2018})}\BibitemShut {NoStop}%
\bibitem [{\citenamefont {Zhou}\ \emph {et~al.}(2021)\citenamefont {Zhou}, \citenamefont {Zhang}, \citenamefont {Yin}, \citenamefont {Huai}, \citenamefont {Gu}, \citenamefont {Xu}, \citenamefont {Allcock}, \citenamefont {Liu}, \citenamefont {Xi}, \citenamefont {Yu} \emph {et~al.}}]{zhou2021rapid}%
  \BibitemOpen
  \bibfield  {author} {\bibinfo {author} {\bibfnamefont {Y.}~\bibnamefont {Zhou}}, \bibinfo {author} {\bibfnamefont {Z.}~\bibnamefont {Zhang}}, \bibinfo {author} {\bibfnamefont {Z.}~\bibnamefont {Yin}}, \bibinfo {author} {\bibfnamefont {S.}~\bibnamefont {Huai}}, \bibinfo {author} {\bibfnamefont {X.}~\bibnamefont {Gu}}, \bibinfo {author} {\bibfnamefont {X.}~\bibnamefont {Xu}}, \bibinfo {author} {\bibfnamefont {J.}~\bibnamefont {Allcock}}, \bibinfo {author} {\bibfnamefont {F.}~\bibnamefont {Liu}}, \bibinfo {author} {\bibfnamefont {G.}~\bibnamefont {Xi}}, \bibinfo {author} {\bibfnamefont {Q.}~\bibnamefont {Yu}}, \emph {et~al.},\ }\bibfield  {title} {\bibinfo {title} {Rapid and unconditional parametric reset protocol for tunable superconducting qubits},\ }\href {https://doi.org/https://doi.org/10.1038/s41467-021-26205-y} {\bibfield  {journal} {\bibinfo  {journal} {Nat. Commun.}\ }\textbf {\bibinfo {volume} {12}},\ \bibinfo {pages} {5924} (\bibinfo {year} {2021})}\BibitemShut {NoStop}%
\bibitem [{\citenamefont {Sunada}\ \emph {et~al.}(2022)\citenamefont {Sunada}, \citenamefont {Kono}, \citenamefont {Ilves}, \citenamefont {Tamate}, \citenamefont {Sugiyama}, \citenamefont {Tabuchi},\ and\ \citenamefont {Nakamura}}]{Sunada2022Fast}%
  \BibitemOpen
  \bibfield  {author} {\bibinfo {author} {\bibfnamefont {Y.}~\bibnamefont {Sunada}}, \bibinfo {author} {\bibfnamefont {S.}~\bibnamefont {Kono}}, \bibinfo {author} {\bibfnamefont {J.}~\bibnamefont {Ilves}}, \bibinfo {author} {\bibfnamefont {S.}~\bibnamefont {Tamate}}, \bibinfo {author} {\bibfnamefont {T.}~\bibnamefont {Sugiyama}}, \bibinfo {author} {\bibfnamefont {Y.}~\bibnamefont {Tabuchi}},\ and\ \bibinfo {author} {\bibfnamefont {Y.}~\bibnamefont {Nakamura}},\ }\bibfield  {title} {\bibinfo {title} {Fast readout and reset of a superconducting qubit coupled to a resonator with an intrinsic purcell filter},\ }\href {https://doi.org/10.1103/PhysRevApplied.17.044016} {\bibfield  {journal} {\bibinfo  {journal} {Phys. Rev. Appl.}\ }\textbf {\bibinfo {volume} {17}},\ \bibinfo {pages} {044016} (\bibinfo {year} {2022})}\BibitemShut {NoStop}%
\bibitem [{\citenamefont {Yoshioka}\ \emph {et~al.}(2023)\citenamefont {Yoshioka}, \citenamefont {Mukai}, \citenamefont {Tomonaga}, \citenamefont {Takada}, \citenamefont {Okazaki}, \citenamefont {Kaneko}, \citenamefont {Nakamura},\ and\ \citenamefont {Tsai}}]{Yoshioka2023Active}%
  \BibitemOpen
  \bibfield  {author} {\bibinfo {author} {\bibfnamefont {T.}~\bibnamefont {Yoshioka}}, \bibinfo {author} {\bibfnamefont {H.}~\bibnamefont {Mukai}}, \bibinfo {author} {\bibfnamefont {A.}~\bibnamefont {Tomonaga}}, \bibinfo {author} {\bibfnamefont {S.}~\bibnamefont {Takada}}, \bibinfo {author} {\bibfnamefont {Y.}~\bibnamefont {Okazaki}}, \bibinfo {author} {\bibfnamefont {N.-H.}\ \bibnamefont {Kaneko}}, \bibinfo {author} {\bibfnamefont {S.}~\bibnamefont {Nakamura}},\ and\ \bibinfo {author} {\bibfnamefont {J.-S.}\ \bibnamefont {Tsai}},\ }\bibfield  {title} {\bibinfo {title} {Active initialization experiment of a superconducting qubit using a quantum circuit refrigerator},\ }\href {https://doi.org/10.1103/PhysRevApplied.20.044077} {\bibfield  {journal} {\bibinfo  {journal} {Phys. Rev. Appl.}\ }\textbf {\bibinfo {volume} {20}},\ \bibinfo {pages} {044077} (\bibinfo {year} {2023})}\BibitemShut {NoStop}%
\bibitem [{\citenamefont {Maurya}\ \emph {et~al.}(2024)\citenamefont {Maurya}, \citenamefont {Zhang}, \citenamefont {Kowsari}, \citenamefont {Kuo}, \citenamefont {Hartsell}, \citenamefont {Miyamoto}, \citenamefont {Liu}, \citenamefont {Shanto}, \citenamefont {Vlachos}, \citenamefont {Zarassi}, \citenamefont {Murch},\ and\ \citenamefont {Levenson-Falk}}]{maurya2023demand}%
  \BibitemOpen
  \bibfield  {author} {\bibinfo {author} {\bibfnamefont {V.}~\bibnamefont {Maurya}}, \bibinfo {author} {\bibfnamefont {H.}~\bibnamefont {Zhang}}, \bibinfo {author} {\bibfnamefont {D.}~\bibnamefont {Kowsari}}, \bibinfo {author} {\bibfnamefont {A.}~\bibnamefont {Kuo}}, \bibinfo {author} {\bibfnamefont {D.~M.}\ \bibnamefont {Hartsell}}, \bibinfo {author} {\bibfnamefont {C.}~\bibnamefont {Miyamoto}}, \bibinfo {author} {\bibfnamefont {J.}~\bibnamefont {Liu}}, \bibinfo {author} {\bibfnamefont {S.}~\bibnamefont {Shanto}}, \bibinfo {author} {\bibfnamefont {E.}~\bibnamefont {Vlachos}}, \bibinfo {author} {\bibfnamefont {A.}~\bibnamefont {Zarassi}}, \bibinfo {author} {\bibfnamefont {K.~W.}\ \bibnamefont {Murch}},\ and\ \bibinfo {author} {\bibfnamefont {E.~M.}\ \bibnamefont {Levenson-Falk}},\ }\bibfield  {title} {\bibinfo {title} {On-demand driven dissipation for cavity reset and cooling},\ }\href {https://link.aps.org/doi/10.1103/PRXQuantum.5.020321} {\bibfield  {journal} {\bibinfo  {journal} {PRX Quantum}\ }\textbf
  {\bibinfo {volume} {5}},\ \bibinfo {pages} {020321} (\bibinfo {year} {2024})}\BibitemShut {NoStop}%
\bibitem [{\citenamefont {Wang}\ \emph {et~al.}(2024)\citenamefont {Wang}, \citenamefont {Wu}, \citenamefont {Wang}, \citenamefont {Ma}, \citenamefont {Zhang}, \citenamefont {Chen}, \citenamefont {Deng}, \citenamefont {Gao}, \citenamefont {Hu}, \citenamefont {Ma}, \citenamefont {Song}, \citenamefont {Xia}, \citenamefont {Ying}, \citenamefont {Zhan}, \citenamefont {Zhao},\ and\ \citenamefont {Deng}}]{Wang2024Efficient}%
  \BibitemOpen
  \bibfield  {author} {\bibinfo {author} {\bibfnamefont {T.}~\bibnamefont {Wang}}, \bibinfo {author} {\bibfnamefont {F.}~\bibnamefont {Wu}}, \bibinfo {author} {\bibfnamefont {F.}~\bibnamefont {Wang}}, \bibinfo {author} {\bibfnamefont {X.}~\bibnamefont {Ma}}, \bibinfo {author} {\bibfnamefont {G.}~\bibnamefont {Zhang}}, \bibinfo {author} {\bibfnamefont {J.}~\bibnamefont {Chen}}, \bibinfo {author} {\bibfnamefont {H.}~\bibnamefont {Deng}}, \bibinfo {author} {\bibfnamefont {R.}~\bibnamefont {Gao}}, \bibinfo {author} {\bibfnamefont {R.}~\bibnamefont {Hu}}, \bibinfo {author} {\bibfnamefont {L.}~\bibnamefont {Ma}}, \bibinfo {author} {\bibfnamefont {Z.}~\bibnamefont {Song}}, \bibinfo {author} {\bibfnamefont {T.}~\bibnamefont {Xia}}, \bibinfo {author} {\bibfnamefont {M.}~\bibnamefont {Ying}}, \bibinfo {author} {\bibfnamefont {H.}~\bibnamefont {Zhan}}, \bibinfo {author} {\bibfnamefont {H.-H.}\ \bibnamefont {Zhao}},\ and\ \bibinfo {author} {\bibfnamefont {C.}~\bibnamefont {Deng}},\ }\bibfield  {title} {\bibinfo {title}
  {Efficient initialization of fluxonium qubits based on auxiliary energy levels},\ }\href {https://link.aps.org/doi/10.1103/PhysRevLett.132.230601} {\bibfield  {journal} {\bibinfo  {journal} {Phys. Rev. Lett.}\ }\textbf {\bibinfo {volume} {132}},\ \bibinfo {pages} {230601} (\bibinfo {year} {2024})}\BibitemShut {NoStop}%
\bibitem [{\citenamefont {McEwen}\ \emph {et~al.}(2021)\citenamefont {McEwen}, \citenamefont {Kafri}, \citenamefont {Chen}, \citenamefont {Atalaya}, \citenamefont {Satzinger}, \citenamefont {Quintana}, \citenamefont {Klimov}, \citenamefont {Sank}, \citenamefont {Gidney}, \citenamefont {Fowler} \emph {et~al.}}]{mcewen2021removing}%
  \BibitemOpen
  \bibfield  {author} {\bibinfo {author} {\bibfnamefont {M.}~\bibnamefont {McEwen}}, \bibinfo {author} {\bibfnamefont {D.}~\bibnamefont {Kafri}}, \bibinfo {author} {\bibfnamefont {Z.}~\bibnamefont {Chen}}, \bibinfo {author} {\bibfnamefont {J.}~\bibnamefont {Atalaya}}, \bibinfo {author} {\bibfnamefont {K.}~\bibnamefont {Satzinger}}, \bibinfo {author} {\bibfnamefont {C.}~\bibnamefont {Quintana}}, \bibinfo {author} {\bibfnamefont {P.~V.}\ \bibnamefont {Klimov}}, \bibinfo {author} {\bibfnamefont {D.}~\bibnamefont {Sank}}, \bibinfo {author} {\bibfnamefont {C.}~\bibnamefont {Gidney}}, \bibinfo {author} {\bibfnamefont {A.}~\bibnamefont {Fowler}}, \emph {et~al.},\ }\bibfield  {title} {\bibinfo {title} {Removing leakage-induced correlated errors in superconducting quantum error correction},\ }\href {https://doi.org/https://doi.org/10.1038/s41467-021-21982-y} {\bibfield  {journal} {\bibinfo  {journal} {Nat. Commun.}\ }\textbf {\bibinfo {volume} {12}},\ \bibinfo {pages} {1761} (\bibinfo {year} {2021})}\BibitemShut
  {NoStop}%
\bibitem [{\citenamefont {Battistel}\ \emph {et~al.}(2021)\citenamefont {Battistel}, \citenamefont {Varbanov},\ and\ \citenamefont {Terhal}}]{Battistel2021Hardware}%
  \BibitemOpen
  \bibfield  {author} {\bibinfo {author} {\bibfnamefont {F.}~\bibnamefont {Battistel}}, \bibinfo {author} {\bibfnamefont {B.}~\bibnamefont {Varbanov}},\ and\ \bibinfo {author} {\bibfnamefont {B.}~\bibnamefont {Terhal}},\ }\bibfield  {title} {\bibinfo {title} {Hardware-efficient leakage-reduction scheme for quantum error correction with superconducting transmon qubits},\ }\href {https://doi.org/10.1103/PRXQuantum.2.030314} {\bibfield  {journal} {\bibinfo  {journal} {PRX Quantum}\ }\textbf {\bibinfo {volume} {2}},\ \bibinfo {pages} {030314} (\bibinfo {year} {2021})}\BibitemShut {NoStop}%
\bibitem [{\citenamefont {Miao}\ \emph {et~al.}(2023)\citenamefont {Miao}, \citenamefont {McEwen}, \citenamefont {Atalaya}, \citenamefont {Kafri}, \citenamefont {Pryadko}, \citenamefont {Bengtsson}, \citenamefont {Opremcak}, \citenamefont {Satzinger}, \citenamefont {Chen}, \citenamefont {Klimov} \emph {et~al.}}]{miao2023overcoming}%
  \BibitemOpen
  \bibfield  {author} {\bibinfo {author} {\bibfnamefont {K.~C.}\ \bibnamefont {Miao}}, \bibinfo {author} {\bibfnamefont {M.}~\bibnamefont {McEwen}}, \bibinfo {author} {\bibfnamefont {J.}~\bibnamefont {Atalaya}}, \bibinfo {author} {\bibfnamefont {D.}~\bibnamefont {Kafri}}, \bibinfo {author} {\bibfnamefont {L.~P.}\ \bibnamefont {Pryadko}}, \bibinfo {author} {\bibfnamefont {A.}~\bibnamefont {Bengtsson}}, \bibinfo {author} {\bibfnamefont {A.}~\bibnamefont {Opremcak}}, \bibinfo {author} {\bibfnamefont {K.~J.}\ \bibnamefont {Satzinger}}, \bibinfo {author} {\bibfnamefont {Z.}~\bibnamefont {Chen}}, \bibinfo {author} {\bibfnamefont {P.~V.}\ \bibnamefont {Klimov}}, \emph {et~al.},\ }\bibfield  {title} {\bibinfo {title} {Overcoming leakage in quantum error correction},\ }\href {https://doi.org/https://doi.org/10.1038/s41567-023-02226-w} {\bibfield  {journal} {\bibinfo  {journal} {Nat. Phys.}\ }\textbf {\bibinfo {volume} {19}},\ \bibinfo {pages} {1780} (\bibinfo {year} {2023})}\BibitemShut {NoStop}%
\bibitem [{\citenamefont {Marques}\ \emph {et~al.}(2023)\citenamefont {Marques}, \citenamefont {Ali}, \citenamefont {Varbanov}, \citenamefont {Finkel}, \citenamefont {Veen}, \citenamefont {van~der Meer}, \citenamefont {Valles-Sanclemente}, \citenamefont {Muthusubramanian}, \citenamefont {Beekman}, \citenamefont {Haider}, \citenamefont {Terhal},\ and\ \citenamefont {DiCarlo}}]{Marques2023All}%
  \BibitemOpen
  \bibfield  {author} {\bibinfo {author} {\bibfnamefont {J.~F.}\ \bibnamefont {Marques}}, \bibinfo {author} {\bibfnamefont {H.}~\bibnamefont {Ali}}, \bibinfo {author} {\bibfnamefont {B.~M.}\ \bibnamefont {Varbanov}}, \bibinfo {author} {\bibfnamefont {M.}~\bibnamefont {Finkel}}, \bibinfo {author} {\bibfnamefont {H.~M.}\ \bibnamefont {Veen}}, \bibinfo {author} {\bibfnamefont {S.~L.~M.}\ \bibnamefont {van~der Meer}}, \bibinfo {author} {\bibfnamefont {S.}~\bibnamefont {Valles-Sanclemente}}, \bibinfo {author} {\bibfnamefont {N.}~\bibnamefont {Muthusubramanian}}, \bibinfo {author} {\bibfnamefont {M.}~\bibnamefont {Beekman}}, \bibinfo {author} {\bibfnamefont {N.}~\bibnamefont {Haider}}, \bibinfo {author} {\bibfnamefont {B.~M.}\ \bibnamefont {Terhal}},\ and\ \bibinfo {author} {\bibfnamefont {L.}~\bibnamefont {DiCarlo}},\ }\bibfield  {title} {\bibinfo {title} {All-microwave leakage reduction units for quantum error correction with superconducting transmon qubits},\ }\href
  {https://doi.org/10.1103/PhysRevLett.130.250602} {\bibfield  {journal} {\bibinfo  {journal} {Phys. Rev. Lett.}\ }\textbf {\bibinfo {volume} {130}},\ \bibinfo {pages} {250602} (\bibinfo {year} {2023})}\BibitemShut {NoStop}%
\bibitem [{\citenamefont {Lacroix}\ \emph {et~al.}(2023)\citenamefont {Lacroix}, \citenamefont {Hofele}, \citenamefont {Remm}, \citenamefont {Benhayoune-Khadraoui}, \citenamefont {McDonald}, \citenamefont {Shillito}, \citenamefont {Lazar}, \citenamefont {Hellings}, \citenamefont {Swiadek}, \citenamefont {Colao-Zanuz} \emph {et~al.}}]{lacroix2023fast}%
  \BibitemOpen
  \bibfield  {author} {\bibinfo {author} {\bibfnamefont {N.}~\bibnamefont {Lacroix}}, \bibinfo {author} {\bibfnamefont {L.}~\bibnamefont {Hofele}}, \bibinfo {author} {\bibfnamefont {A.}~\bibnamefont {Remm}}, \bibinfo {author} {\bibfnamefont {O.}~\bibnamefont {Benhayoune-Khadraoui}}, \bibinfo {author} {\bibfnamefont {A.}~\bibnamefont {McDonald}}, \bibinfo {author} {\bibfnamefont {R.}~\bibnamefont {Shillito}}, \bibinfo {author} {\bibfnamefont {S.}~\bibnamefont {Lazar}}, \bibinfo {author} {\bibfnamefont {C.}~\bibnamefont {Hellings}}, \bibinfo {author} {\bibfnamefont {F.}~\bibnamefont {Swiadek}}, \bibinfo {author} {\bibfnamefont {D.}~\bibnamefont {Colao-Zanuz}}, \emph {et~al.},\ }\bibfield  {title} {\bibinfo {title} {Fast flux-activated leakage reduction for superconducting quantum circuits},\ }\href {https://doi.org/10.48550/arXiv.2309.07060} {\bibfield  {journal} {\bibinfo  {journal} {arXiv:2309.07060}\ } (\bibinfo {year} {2023})}\BibitemShut {NoStop}%
\bibitem [{\citenamefont {Yang}\ \emph {et~al.}(2024)\citenamefont {Yang}, \citenamefont {Chu}, \citenamefont {Guo}, \citenamefont {Huang}, \citenamefont {Liang}, \citenamefont {Liu}, \citenamefont {Qiu}, \citenamefont {Sun}, \citenamefont {Tao}, \citenamefont {Zhang}, \citenamefont {Zhang}, \citenamefont {Zhang}, \citenamefont {Zhou}, \citenamefont {Guo}, \citenamefont {Hu}, \citenamefont {Jiang}, \citenamefont {Liu}, \citenamefont {Linpeng}, \citenamefont {Chen}, \citenamefont {Chen}, \citenamefont {Niu}, \citenamefont {Liu}, \citenamefont {Zhong},\ and\ \citenamefont {Yu}}]{yang2024coupler}%
  \BibitemOpen
  \bibfield  {author} {\bibinfo {author} {\bibfnamefont {X.}~\bibnamefont {Yang}}, \bibinfo {author} {\bibfnamefont {J.}~\bibnamefont {Chu}}, \bibinfo {author} {\bibfnamefont {Z.}~\bibnamefont {Guo}}, \bibinfo {author} {\bibfnamefont {W.}~\bibnamefont {Huang}}, \bibinfo {author} {\bibfnamefont {Y.}~\bibnamefont {Liang}}, \bibinfo {author} {\bibfnamefont {J.}~\bibnamefont {Liu}}, \bibinfo {author} {\bibfnamefont {J.}~\bibnamefont {Qiu}}, \bibinfo {author} {\bibfnamefont {X.}~\bibnamefont {Sun}}, \bibinfo {author} {\bibfnamefont {Z.}~\bibnamefont {Tao}}, \bibinfo {author} {\bibfnamefont {J.}~\bibnamefont {Zhang}}, \bibinfo {author} {\bibfnamefont {J.}~\bibnamefont {Zhang}}, \bibinfo {author} {\bibfnamefont {L.}~\bibnamefont {Zhang}}, \bibinfo {author} {\bibfnamefont {Y.}~\bibnamefont {Zhou}}, \bibinfo {author} {\bibfnamefont {W.}~\bibnamefont {Guo}}, \bibinfo {author} {\bibfnamefont {L.}~\bibnamefont {Hu}}, \bibinfo {author} {\bibfnamefont {J.}~\bibnamefont {Jiang}}, \bibinfo {author} {\bibfnamefont
  {Y.}~\bibnamefont {Liu}}, \bibinfo {author} {\bibfnamefont {X.}~\bibnamefont {Linpeng}}, \bibinfo {author} {\bibfnamefont {T.}~\bibnamefont {Chen}}, \bibinfo {author} {\bibfnamefont {Y.}~\bibnamefont {Chen}}, \bibinfo {author} {\bibfnamefont {J.}~\bibnamefont {Niu}}, \bibinfo {author} {\bibfnamefont {S.}~\bibnamefont {Liu}}, \bibinfo {author} {\bibfnamefont {Y.}~\bibnamefont {Zhong}},\ and\ \bibinfo {author} {\bibfnamefont {D.}~\bibnamefont {Yu}},\ }\bibfield  {title} {\bibinfo {title} {Coupler-assisted leakage reduction for scalable quantum error correction with superconducting qubits},\ }\href {https://doi.org/10.1103/PhysRevLett.133.170601} {\bibfield  {journal} {\bibinfo  {journal} {Phys. Rev. Lett.}\ }\textbf {\bibinfo {volume} {133}},\ \bibinfo {pages} {170601} (\bibinfo {year} {2024})}\BibitemShut {NoStop}%
\bibitem [{\citenamefont {Mi}\ \emph {et~al.}(2024)\citenamefont {Mi}, \citenamefont {Michailidis}, \citenamefont {Shabani}, \citenamefont {Miao}, \citenamefont {Klimov}, \citenamefont {Lloyd}, \citenamefont {Rosenberg}, \citenamefont {Acharya}, \citenamefont {Aleiner}, \citenamefont {Andersen} \emph {et~al.}}]{mi2024stable}%
  \BibitemOpen
  \bibfield  {author} {\bibinfo {author} {\bibfnamefont {X.}~\bibnamefont {Mi}}, \bibinfo {author} {\bibfnamefont {A.}~\bibnamefont {Michailidis}}, \bibinfo {author} {\bibfnamefont {S.}~\bibnamefont {Shabani}}, \bibinfo {author} {\bibfnamefont {K.}~\bibnamefont {Miao}}, \bibinfo {author} {\bibfnamefont {P.}~\bibnamefont {Klimov}}, \bibinfo {author} {\bibfnamefont {J.}~\bibnamefont {Lloyd}}, \bibinfo {author} {\bibfnamefont {E.}~\bibnamefont {Rosenberg}}, \bibinfo {author} {\bibfnamefont {R.}~\bibnamefont {Acharya}}, \bibinfo {author} {\bibfnamefont {I.}~\bibnamefont {Aleiner}}, \bibinfo {author} {\bibfnamefont {T.}~\bibnamefont {Andersen}}, \emph {et~al.},\ }\bibfield  {title} {\bibinfo {title} {Stable quantum-correlated many-body states through engineered dissipation},\ }\href {https://www.science.org/doi/10.1126/science.adh9932} {\bibfield  {journal} {\bibinfo  {journal} {Science}\ }\textbf {\bibinfo {volume} {383}},\ \bibinfo {pages} {1332} (\bibinfo {year} {2024})}\BibitemShut {NoStop}%
\bibitem [{\citenamefont {DeCross}\ \emph {et~al.}(2023)\citenamefont {DeCross}, \citenamefont {Chertkov}, \citenamefont {Kohagen},\ and\ \citenamefont {Foss-Feig}}]{DeCross2023Qubit}%
  \BibitemOpen
  \bibfield  {author} {\bibinfo {author} {\bibfnamefont {M.}~\bibnamefont {DeCross}}, \bibinfo {author} {\bibfnamefont {E.}~\bibnamefont {Chertkov}}, \bibinfo {author} {\bibfnamefont {M.}~\bibnamefont {Kohagen}},\ and\ \bibinfo {author} {\bibfnamefont {M.}~\bibnamefont {Foss-Feig}},\ }\bibfield  {title} {\bibinfo {title} {Qubit-reuse compilation with mid-circuit measurement and reset},\ }\href {https://doi.org/10.1103/PhysRevX.13.041057} {\bibfield  {journal} {\bibinfo  {journal} {Phys. Rev. X}\ }\textbf {\bibinfo {volume} {13}},\ \bibinfo {pages} {041057} (\bibinfo {year} {2023})}\BibitemShut {NoStop}%
\bibitem [{\citenamefont {Barreiro}\ \emph {et~al.}(2011)\citenamefont {Barreiro}, \citenamefont {M{\"u}ller}, \citenamefont {Schindler}, \citenamefont {Nigg}, \citenamefont {Monz}, \citenamefont {Chwalla}, \citenamefont {Hennrich}, \citenamefont {Roos}, \citenamefont {Zoller},\ and\ \citenamefont {Blatt}}]{barreiro2011open}%
  \BibitemOpen
  \bibfield  {author} {\bibinfo {author} {\bibfnamefont {J.~T.}\ \bibnamefont {Barreiro}}, \bibinfo {author} {\bibfnamefont {M.}~\bibnamefont {M{\"u}ller}}, \bibinfo {author} {\bibfnamefont {P.}~\bibnamefont {Schindler}}, \bibinfo {author} {\bibfnamefont {D.}~\bibnamefont {Nigg}}, \bibinfo {author} {\bibfnamefont {T.}~\bibnamefont {Monz}}, \bibinfo {author} {\bibfnamefont {M.}~\bibnamefont {Chwalla}}, \bibinfo {author} {\bibfnamefont {M.}~\bibnamefont {Hennrich}}, \bibinfo {author} {\bibfnamefont {C.~F.}\ \bibnamefont {Roos}}, \bibinfo {author} {\bibfnamefont {P.}~\bibnamefont {Zoller}},\ and\ \bibinfo {author} {\bibfnamefont {R.}~\bibnamefont {Blatt}},\ }\bibfield  {title} {\bibinfo {title} {An open-system quantum simulator with trapped ions},\ }\href {https://doi.org/https://doi.org/10.1038/nature09801} {\bibfield  {journal} {\bibinfo  {journal} {Nature}\ }\textbf {\bibinfo {volume} {470}},\ \bibinfo {pages} {486} (\bibinfo {year} {2011})}\BibitemShut {NoStop}%
\bibitem [{\citenamefont {Kapit}\ \emph {et~al.}(2020)\citenamefont {Kapit}, \citenamefont {Roushan}, \citenamefont {Neill}, \citenamefont {Boixo},\ and\ \citenamefont {Smelyanskiy}}]{kapit2020entanglement}%
  \BibitemOpen
  \bibfield  {author} {\bibinfo {author} {\bibfnamefont {E.}~\bibnamefont {Kapit}}, \bibinfo {author} {\bibfnamefont {P.}~\bibnamefont {Roushan}}, \bibinfo {author} {\bibfnamefont {C.}~\bibnamefont {Neill}}, \bibinfo {author} {\bibfnamefont {S.}~\bibnamefont {Boixo}},\ and\ \bibinfo {author} {\bibfnamefont {V.}~\bibnamefont {Smelyanskiy}},\ }\bibfield  {title} {\bibinfo {title} {Entanglement and complexity of interacting qubits subject to asymmetric noise},\ }\href {https://doi.org/10.1103/PhysRevResearch.2.043042} {\bibfield  {journal} {\bibinfo  {journal} {Phys. Rev. Res.}\ }\textbf {\bibinfo {volume} {2}},\ \bibinfo {pages} {043042} (\bibinfo {year} {2020})}\BibitemShut {NoStop}%
\bibitem [{\citenamefont {Rist\`e}\ \emph {et~al.}(2012)\citenamefont {Rist\`e}, \citenamefont {van Leeuwen}, \citenamefont {Ku}, \citenamefont {Lehnert},\ and\ \citenamefont {DiCarlo}}]{2012Initialization}%
  \BibitemOpen
  \bibfield  {author} {\bibinfo {author} {\bibfnamefont {D.}~\bibnamefont {Rist\`e}}, \bibinfo {author} {\bibfnamefont {J.~G.}\ \bibnamefont {van Leeuwen}}, \bibinfo {author} {\bibfnamefont {H.-S.}\ \bibnamefont {Ku}}, \bibinfo {author} {\bibfnamefont {K.~W.}\ \bibnamefont {Lehnert}},\ and\ \bibinfo {author} {\bibfnamefont {L.}~\bibnamefont {DiCarlo}},\ }\bibfield  {title} {\bibinfo {title} {Initialization by measurement of a superconducting quantum bit circuit},\ }\href {https://doi.org/10.1103/PhysRevLett.109.050507} {\bibfield  {journal} {\bibinfo  {journal} {Phys. Rev. Lett.}\ }\textbf {\bibinfo {volume} {109}},\ \bibinfo {pages} {050507} (\bibinfo {year} {2012})}\BibitemShut {NoStop}%
\bibitem [{\citenamefont {Salath\'e}\ \emph {et~al.}(2018)\citenamefont {Salath\'e}, \citenamefont {Kurpiers}, \citenamefont {Karg}, \citenamefont {Lang}, \citenamefont {Andersen}, \citenamefont {Akin}, \citenamefont {Krinner}, \citenamefont {Eichler},\ and\ \citenamefont {Wallraff}}]{salathe2018low}%
  \BibitemOpen
  \bibfield  {author} {\bibinfo {author} {\bibfnamefont {Y.}~\bibnamefont {Salath\'e}}, \bibinfo {author} {\bibfnamefont {P.}~\bibnamefont {Kurpiers}}, \bibinfo {author} {\bibfnamefont {T.}~\bibnamefont {Karg}}, \bibinfo {author} {\bibfnamefont {C.}~\bibnamefont {Lang}}, \bibinfo {author} {\bibfnamefont {C.~K.}\ \bibnamefont {Andersen}}, \bibinfo {author} {\bibfnamefont {A.}~\bibnamefont {Akin}}, \bibinfo {author} {\bibfnamefont {S.}~\bibnamefont {Krinner}}, \bibinfo {author} {\bibfnamefont {C.}~\bibnamefont {Eichler}},\ and\ \bibinfo {author} {\bibfnamefont {A.}~\bibnamefont {Wallraff}},\ }\bibfield  {title} {\bibinfo {title} {Low-latency digital signal processing for feedback and feedforward in quantum computing and communication},\ }\href {https://doi.org/10.1103/PhysRevApplied.9.034011} {\bibfield  {journal} {\bibinfo  {journal} {Phys. Rev. Appl.}\ }\textbf {\bibinfo {volume} {9}},\ \bibinfo {pages} {034011} (\bibinfo {year} {2018})}\BibitemShut {NoStop}%
\bibitem [{\citenamefont {Jeffrey}\ \emph {et~al.}(2014)\citenamefont {Jeffrey}, \citenamefont {Sank}, \citenamefont {Mutus}, \citenamefont {White}, \citenamefont {Kelly}, \citenamefont {Barends}, \citenamefont {Chen}, \citenamefont {Chen}, \citenamefont {Chiaro}, \citenamefont {Dunsworth}, \citenamefont {Megrant}, \citenamefont {O'Malley}, \citenamefont {Neill}, \citenamefont {Roushan}, \citenamefont {Vainsencher}, \citenamefont {Wenner}, \citenamefont {Cleland},\ and\ \citenamefont {Martinis}}]{jeffrey2014Fast}%
  \BibitemOpen
  \bibfield  {author} {\bibinfo {author} {\bibfnamefont {E.}~\bibnamefont {Jeffrey}}, \bibinfo {author} {\bibfnamefont {D.}~\bibnamefont {Sank}}, \bibinfo {author} {\bibfnamefont {J.~Y.}\ \bibnamefont {Mutus}}, \bibinfo {author} {\bibfnamefont {T.~C.}\ \bibnamefont {White}}, \bibinfo {author} {\bibfnamefont {J.}~\bibnamefont {Kelly}}, \bibinfo {author} {\bibfnamefont {R.}~\bibnamefont {Barends}}, \bibinfo {author} {\bibfnamefont {Y.}~\bibnamefont {Chen}}, \bibinfo {author} {\bibfnamefont {Z.}~\bibnamefont {Chen}}, \bibinfo {author} {\bibfnamefont {B.}~\bibnamefont {Chiaro}}, \bibinfo {author} {\bibfnamefont {A.}~\bibnamefont {Dunsworth}}, \bibinfo {author} {\bibfnamefont {A.}~\bibnamefont {Megrant}}, \bibinfo {author} {\bibfnamefont {P.~J.~J.}\ \bibnamefont {O'Malley}}, \bibinfo {author} {\bibfnamefont {C.}~\bibnamefont {Neill}}, \bibinfo {author} {\bibfnamefont {P.}~\bibnamefont {Roushan}}, \bibinfo {author} {\bibfnamefont {A.}~\bibnamefont {Vainsencher}}, \bibinfo {author} {\bibfnamefont {J.}~\bibnamefont
  {Wenner}}, \bibinfo {author} {\bibfnamefont {A.~N.}\ \bibnamefont {Cleland}},\ and\ \bibinfo {author} {\bibfnamefont {J.~M.}\ \bibnamefont {Martinis}},\ }\bibfield  {title} {\bibinfo {title} {Fast accurate state measurement with superconducting qubits},\ }\href {https://link.aps.org/doi/10.1103/PhysRevLett.112.190504} {\bibfield  {journal} {\bibinfo  {journal} {Phys. Rev. Lett.}\ }\textbf {\bibinfo {volume} {112}},\ \bibinfo {pages} {190504} (\bibinfo {year} {2014})}\BibitemShut {NoStop}%
\bibitem [{\citenamefont {Walter}\ \emph {et~al.}(2017)\citenamefont {Walter}, \citenamefont {Kurpiers}, \citenamefont {Gasparinetti}, \citenamefont {Magnard}, \citenamefont {Poto\ifmmode~\check{c}\else \v{c}\fi{}nik}, \citenamefont {Salath\'e}, \citenamefont {Pechal}, \citenamefont {Mondal}, \citenamefont {Oppliger}, \citenamefont {Eichler},\ and\ \citenamefont {Wallraff}}]{Walter2017Rapid}%
  \BibitemOpen
  \bibfield  {author} {\bibinfo {author} {\bibfnamefont {T.}~\bibnamefont {Walter}}, \bibinfo {author} {\bibfnamefont {P.}~\bibnamefont {Kurpiers}}, \bibinfo {author} {\bibfnamefont {S.}~\bibnamefont {Gasparinetti}}, \bibinfo {author} {\bibfnamefont {P.}~\bibnamefont {Magnard}}, \bibinfo {author} {\bibfnamefont {A.}~\bibnamefont {Poto\ifmmode~\check{c}\else \v{c}\fi{}nik}}, \bibinfo {author} {\bibfnamefont {Y.}~\bibnamefont {Salath\'e}}, \bibinfo {author} {\bibfnamefont {M.}~\bibnamefont {Pechal}}, \bibinfo {author} {\bibfnamefont {M.}~\bibnamefont {Mondal}}, \bibinfo {author} {\bibfnamefont {M.}~\bibnamefont {Oppliger}}, \bibinfo {author} {\bibfnamefont {C.}~\bibnamefont {Eichler}},\ and\ \bibinfo {author} {\bibfnamefont {A.}~\bibnamefont {Wallraff}},\ }\bibfield  {title} {\bibinfo {title} {Rapid high-fidelity single-shot dispersive readout of superconducting qubits},\ }\href {https://link.aps.org/doi/10.1103/PhysRevApplied.7.054020} {\bibfield  {journal} {\bibinfo  {journal} {Phys. Rev. Appl.}\ }\textbf
  {\bibinfo {volume} {7}},\ \bibinfo {pages} {054020} (\bibinfo {year} {2017})}\BibitemShut {NoStop}%
\bibitem [{\citenamefont {Sunada}\ \emph {et~al.}(2024)\citenamefont {Sunada}, \citenamefont {Yuki}, \citenamefont {Wang}, \citenamefont {Miyamura}, \citenamefont {Ilves}, \citenamefont {Matsuura}, \citenamefont {Spring}, \citenamefont {Tamate}, \citenamefont {Kono},\ and\ \citenamefont {Nakamura}}]{Sunada2024Photon}%
  \BibitemOpen
  \bibfield  {author} {\bibinfo {author} {\bibfnamefont {Y.}~\bibnamefont {Sunada}}, \bibinfo {author} {\bibfnamefont {K.}~\bibnamefont {Yuki}}, \bibinfo {author} {\bibfnamefont {Z.}~\bibnamefont {Wang}}, \bibinfo {author} {\bibfnamefont {T.}~\bibnamefont {Miyamura}}, \bibinfo {author} {\bibfnamefont {J.}~\bibnamefont {Ilves}}, \bibinfo {author} {\bibfnamefont {K.}~\bibnamefont {Matsuura}}, \bibinfo {author} {\bibfnamefont {P.~A.}\ \bibnamefont {Spring}}, \bibinfo {author} {\bibfnamefont {S.}~\bibnamefont {Tamate}}, \bibinfo {author} {\bibfnamefont {S.}~\bibnamefont {Kono}},\ and\ \bibinfo {author} {\bibfnamefont {Y.}~\bibnamefont {Nakamura}},\ }\bibfield  {title} {\bibinfo {title} {Photon-noise-tolerant dispersive readout of a superconducting qubit using a nonlinear purcell filter},\ }\href {https://link.aps.org/doi/10.1103/PRXQuantum.5.010307} {\bibfield  {journal} {\bibinfo  {journal} {PRX Quantum}\ }\textbf {\bibinfo {volume} {5}},\ \bibinfo {pages} {010307} (\bibinfo {year} {2024})}\BibitemShut {NoStop}%
\bibitem [{\citenamefont {Yen}\ \emph {et~al.}(2024)\citenamefont {Yen}, \citenamefont {Ye}, \citenamefont {Peng}, \citenamefont {Wang}, \citenamefont {Cunningham}, \citenamefont {Gingras}, \citenamefont {Niedzielski}, \citenamefont {Stickler}, \citenamefont {Serniak}, \citenamefont {Schwartz} \emph {et~al.}}]{yen2024interferometric}%
  \BibitemOpen
  \bibfield  {author} {\bibinfo {author} {\bibfnamefont {A.}~\bibnamefont {Yen}}, \bibinfo {author} {\bibfnamefont {Y.}~\bibnamefont {Ye}}, \bibinfo {author} {\bibfnamefont {K.}~\bibnamefont {Peng}}, \bibinfo {author} {\bibfnamefont {J.}~\bibnamefont {Wang}}, \bibinfo {author} {\bibfnamefont {G.}~\bibnamefont {Cunningham}}, \bibinfo {author} {\bibfnamefont {M.}~\bibnamefont {Gingras}}, \bibinfo {author} {\bibfnamefont {B.~M.}\ \bibnamefont {Niedzielski}}, \bibinfo {author} {\bibfnamefont {H.}~\bibnamefont {Stickler}}, \bibinfo {author} {\bibfnamefont {K.}~\bibnamefont {Serniak}}, \bibinfo {author} {\bibfnamefont {M.~E.}\ \bibnamefont {Schwartz}}, \emph {et~al.},\ }\bibfield  {title} {\bibinfo {title} {Interferometric purcell suppression of spontaneous emission in a superconducting qubit},\ }\href {https://arxiv.org/abs/2405.10107} {\bibfield  {journal} {\bibinfo  {journal} {arXiv:2405.10107}\ } (\bibinfo {year} {2024})}\BibitemShut {NoStop}%
\bibitem [{\citenamefont {Sank}\ \emph {et~al.}(2016)\citenamefont {Sank}, \citenamefont {Chen}, \citenamefont {Khezri}, \citenamefont {Kelly}, \citenamefont {Barends}, \citenamefont {Campbell}, \citenamefont {Chen}, \citenamefont {Chiaro}, \citenamefont {Dunsworth}, \citenamefont {Fowler}, \citenamefont {Jeffrey}, \citenamefont {Lucero}, \citenamefont {Megrant}, \citenamefont {Mutus}, \citenamefont {Neeley}, \citenamefont {Neill}, \citenamefont {O'Malley}, \citenamefont {Quintana}, \citenamefont {Roushan}, \citenamefont {Vainsencher}, \citenamefont {White}, \citenamefont {Wenner}, \citenamefont {Korotkov},\ and\ \citenamefont {Martinis}}]{Sank2016Measurement}%
  \BibitemOpen
  \bibfield  {author} {\bibinfo {author} {\bibfnamefont {D.}~\bibnamefont {Sank}}, \bibinfo {author} {\bibfnamefont {Z.}~\bibnamefont {Chen}}, \bibinfo {author} {\bibfnamefont {M.}~\bibnamefont {Khezri}}, \bibinfo {author} {\bibfnamefont {J.}~\bibnamefont {Kelly}}, \bibinfo {author} {\bibfnamefont {R.}~\bibnamefont {Barends}}, \bibinfo {author} {\bibfnamefont {B.}~\bibnamefont {Campbell}}, \bibinfo {author} {\bibfnamefont {Y.}~\bibnamefont {Chen}}, \bibinfo {author} {\bibfnamefont {B.}~\bibnamefont {Chiaro}}, \bibinfo {author} {\bibfnamefont {A.}~\bibnamefont {Dunsworth}}, \bibinfo {author} {\bibfnamefont {A.}~\bibnamefont {Fowler}}, \bibinfo {author} {\bibfnamefont {E.}~\bibnamefont {Jeffrey}}, \bibinfo {author} {\bibfnamefont {E.}~\bibnamefont {Lucero}}, \bibinfo {author} {\bibfnamefont {A.}~\bibnamefont {Megrant}}, \bibinfo {author} {\bibfnamefont {J.}~\bibnamefont {Mutus}}, \bibinfo {author} {\bibfnamefont {M.}~\bibnamefont {Neeley}}, \bibinfo {author} {\bibfnamefont {C.}~\bibnamefont {Neill}}, \bibinfo
  {author} {\bibfnamefont {P.~J.~J.}\ \bibnamefont {O'Malley}}, \bibinfo {author} {\bibfnamefont {C.}~\bibnamefont {Quintana}}, \bibinfo {author} {\bibfnamefont {P.}~\bibnamefont {Roushan}}, \bibinfo {author} {\bibfnamefont {A.}~\bibnamefont {Vainsencher}}, \bibinfo {author} {\bibfnamefont {T.}~\bibnamefont {White}}, \bibinfo {author} {\bibfnamefont {J.}~\bibnamefont {Wenner}}, \bibinfo {author} {\bibfnamefont {A.~N.}\ \bibnamefont {Korotkov}},\ and\ \bibinfo {author} {\bibfnamefont {J.~M.}\ \bibnamefont {Martinis}},\ }\bibfield  {title} {\bibinfo {title} {Measurement-induced state transitions in a superconducting qubit: Beyond the rotating wave approximation},\ }\href {https://doi.org/10.1103/PhysRevLett.117.190503} {\bibfield  {journal} {\bibinfo  {journal} {Phys. Rev. Lett.}\ }\textbf {\bibinfo {volume} {117}},\ \bibinfo {pages} {190503} (\bibinfo {year} {2016})}\BibitemShut {NoStop}%
\bibitem [{\citenamefont {Shillito}\ \emph {et~al.}(2022)\citenamefont {Shillito}, \citenamefont {Petrescu}, \citenamefont {Cohen}, \citenamefont {Beall}, \citenamefont {Hauru}, \citenamefont {Ganahl}, \citenamefont {Lewis}, \citenamefont {Vidal},\ and\ \citenamefont {Blais}}]{shillito2022Dynamics}%
  \BibitemOpen
  \bibfield  {author} {\bibinfo {author} {\bibfnamefont {R.}~\bibnamefont {Shillito}}, \bibinfo {author} {\bibfnamefont {A.}~\bibnamefont {Petrescu}}, \bibinfo {author} {\bibfnamefont {J.}~\bibnamefont {Cohen}}, \bibinfo {author} {\bibfnamefont {J.}~\bibnamefont {Beall}}, \bibinfo {author} {\bibfnamefont {M.}~\bibnamefont {Hauru}}, \bibinfo {author} {\bibfnamefont {M.}~\bibnamefont {Ganahl}}, \bibinfo {author} {\bibfnamefont {A.~G.}\ \bibnamefont {Lewis}}, \bibinfo {author} {\bibfnamefont {G.}~\bibnamefont {Vidal}},\ and\ \bibinfo {author} {\bibfnamefont {A.}~\bibnamefont {Blais}},\ }\bibfield  {title} {\bibinfo {title} {Dynamics of transmon ionization},\ }\href {https://link.aps.org/doi/10.1103/PhysRevApplied.18.034031} {\bibfield  {journal} {\bibinfo  {journal} {Phys. Rev. Appl.}\ }\textbf {\bibinfo {volume} {18}},\ \bibinfo {pages} {034031} (\bibinfo {year} {2022})}\BibitemShut {NoStop}%
\bibitem [{\citenamefont {Khezri}\ \emph {et~al.}(2023)\citenamefont {Khezri}, \citenamefont {Opremcak}, \citenamefont {Chen}, \citenamefont {Miao}, \citenamefont {McEwen}, \citenamefont {Bengtsson}, \citenamefont {White}, \citenamefont {Naaman}, \citenamefont {Sank}, \citenamefont {Korotkov}, \citenamefont {Chen},\ and\ \citenamefont {Smelyanskiy}}]{Khezri2023Measurement}%
  \BibitemOpen
  \bibfield  {author} {\bibinfo {author} {\bibfnamefont {M.}~\bibnamefont {Khezri}}, \bibinfo {author} {\bibfnamefont {A.}~\bibnamefont {Opremcak}}, \bibinfo {author} {\bibfnamefont {Z.}~\bibnamefont {Chen}}, \bibinfo {author} {\bibfnamefont {K.~C.}\ \bibnamefont {Miao}}, \bibinfo {author} {\bibfnamefont {M.}~\bibnamefont {McEwen}}, \bibinfo {author} {\bibfnamefont {A.}~\bibnamefont {Bengtsson}}, \bibinfo {author} {\bibfnamefont {T.}~\bibnamefont {White}}, \bibinfo {author} {\bibfnamefont {O.}~\bibnamefont {Naaman}}, \bibinfo {author} {\bibfnamefont {D.}~\bibnamefont {Sank}}, \bibinfo {author} {\bibfnamefont {A.~N.}\ \bibnamefont {Korotkov}}, \bibinfo {author} {\bibfnamefont {Y.}~\bibnamefont {Chen}},\ and\ \bibinfo {author} {\bibfnamefont {V.}~\bibnamefont {Smelyanskiy}},\ }\bibfield  {title} {\bibinfo {title} {Measurement-induced state transitions in a superconducting qubit: Within the rotating-wave approximation},\ }\href {https://link.aps.org/doi/10.1103/PhysRevApplied.20.054008} {\bibfield  {journal}
  {\bibinfo  {journal} {Phys. Rev. Appl.}\ }\textbf {\bibinfo {volume} {20}},\ \bibinfo {pages} {054008} (\bibinfo {year} {2023})}\BibitemShut {NoStop}%
\bibitem [{\citenamefont {Murch}\ \emph {et~al.}(2012)\citenamefont {Murch}, \citenamefont {Vool}, \citenamefont {Zhou}, \citenamefont {Weber}, \citenamefont {Girvin},\ and\ \citenamefont {Siddiqi}}]{murch2012Cavity}%
  \BibitemOpen
  \bibfield  {author} {\bibinfo {author} {\bibfnamefont {K.~W.}\ \bibnamefont {Murch}}, \bibinfo {author} {\bibfnamefont {U.}~\bibnamefont {Vool}}, \bibinfo {author} {\bibfnamefont {D.}~\bibnamefont {Zhou}}, \bibinfo {author} {\bibfnamefont {S.~J.}\ \bibnamefont {Weber}}, \bibinfo {author} {\bibfnamefont {S.~M.}\ \bibnamefont {Girvin}},\ and\ \bibinfo {author} {\bibfnamefont {I.}~\bibnamefont {Siddiqi}},\ }\bibfield  {title} {\bibinfo {title} {Cavity-assisted quantum bath engineering},\ }\href {https://link.aps.org/doi/10.1103/PhysRevLett.109.183602} {\bibfield  {journal} {\bibinfo  {journal} {Phys. Rev. Lett.}\ }\textbf {\bibinfo {volume} {109}},\ \bibinfo {pages} {183602} (\bibinfo {year} {2012})}\BibitemShut {NoStop}%
\bibitem [{\citenamefont {Yoshioka}\ and\ \citenamefont {Tsai}(2021)}]{yoshioka2021fast}%
  \BibitemOpen
  \bibfield  {author} {\bibinfo {author} {\bibfnamefont {T.}~\bibnamefont {Yoshioka}}\ and\ \bibinfo {author} {\bibfnamefont {J.}~\bibnamefont {Tsai}},\ }\bibfield  {title} {\bibinfo {title} {Fast unconditional initialization for superconducting qubit and resonator using quantum-circuit refrigerator},\ }\href {https://doi.org/10.1063/5.0057894} {\bibfield  {journal} {\bibinfo  {journal} {Appl. Phys. Lett.}\ }\textbf {\bibinfo {volume} {119}} (\bibinfo {year} {2021})}\BibitemShut {NoStop}%
\bibitem [{\citenamefont {Sevriuk}\ \emph {et~al.}(2022)\citenamefont {Sevriuk}, \citenamefont {Liu}, \citenamefont {R{\"o}nkk{\"o}}, \citenamefont {Hsu}, \citenamefont {Marxer}, \citenamefont {M{\"o}rstedt}, \citenamefont {Partanen}, \citenamefont {R{\"a}bin{\"a}}, \citenamefont {Venkatesh}, \citenamefont {Hotari} \emph {et~al.}}]{sevriuk2022initial}%
  \BibitemOpen
  \bibfield  {author} {\bibinfo {author} {\bibfnamefont {V.}~\bibnamefont {Sevriuk}}, \bibinfo {author} {\bibfnamefont {W.}~\bibnamefont {Liu}}, \bibinfo {author} {\bibfnamefont {J.}~\bibnamefont {R{\"o}nkk{\"o}}}, \bibinfo {author} {\bibfnamefont {H.}~\bibnamefont {Hsu}}, \bibinfo {author} {\bibfnamefont {F.}~\bibnamefont {Marxer}}, \bibinfo {author} {\bibfnamefont {T.}~\bibnamefont {M{\"o}rstedt}}, \bibinfo {author} {\bibfnamefont {M.}~\bibnamefont {Partanen}}, \bibinfo {author} {\bibfnamefont {J.}~\bibnamefont {R{\"a}bin{\"a}}}, \bibinfo {author} {\bibfnamefont {M.}~\bibnamefont {Venkatesh}}, \bibinfo {author} {\bibfnamefont {J.}~\bibnamefont {Hotari}}, \emph {et~al.},\ }\bibfield  {title} {\bibinfo {title} {Initial experimental results on a superconducting-qubit reset based on photon-assisted quasiparticle tunneling},\ }\href {https://doi.org/10.1063/5.0129345} {\bibfield  {journal} {\bibinfo  {journal} {Appl. Phys. Lett.}\ }\textbf {\bibinfo {volume} {121}} (\bibinfo {year} {2022})}\BibitemShut {NoStop}%
\bibitem [{\citenamefont {Pan}\ \emph {et~al.}(2022)\citenamefont {Pan}, \citenamefont {Zhou}, \citenamefont {Yuan}, \citenamefont {Nie}, \citenamefont {Wei}, \citenamefont {Zhang}, \citenamefont {Li}, \citenamefont {Liu}, \citenamefont {Jiang}, \citenamefont {Catelani} \emph {et~al.}}]{pan2022engineering}%
  \BibitemOpen
  \bibfield  {author} {\bibinfo {author} {\bibfnamefont {X.}~\bibnamefont {Pan}}, \bibinfo {author} {\bibfnamefont {Y.}~\bibnamefont {Zhou}}, \bibinfo {author} {\bibfnamefont {H.}~\bibnamefont {Yuan}}, \bibinfo {author} {\bibfnamefont {L.}~\bibnamefont {Nie}}, \bibinfo {author} {\bibfnamefont {W.}~\bibnamefont {Wei}}, \bibinfo {author} {\bibfnamefont {L.}~\bibnamefont {Zhang}}, \bibinfo {author} {\bibfnamefont {J.}~\bibnamefont {Li}}, \bibinfo {author} {\bibfnamefont {S.}~\bibnamefont {Liu}}, \bibinfo {author} {\bibfnamefont {Z.~H.}\ \bibnamefont {Jiang}}, \bibinfo {author} {\bibfnamefont {G.}~\bibnamefont {Catelani}}, \emph {et~al.},\ }\bibfield  {title} {\bibinfo {title} {Engineering superconducting qubits to reduce quasiparticles and charge noise},\ }\href {https://www.nature.com/articles/s41467-022-34727-2} {\bibfield  {journal} {\bibinfo  {journal} {Nat. Commun.}\ }\textbf {\bibinfo {volume} {13}},\ \bibinfo {pages} {7196} (\bibinfo {year} {2022})}\BibitemShut {NoStop}%
\bibitem [{\citenamefont {Heinsoo}\ \emph {et~al.}(2018)\citenamefont {Heinsoo}, \citenamefont {Andersen}, \citenamefont {Remm}, \citenamefont {Krinner}, \citenamefont {Walter}, \citenamefont {Salath{\'e}}, \citenamefont {Gasparinetti}, \citenamefont {Besse}, \citenamefont {Poto{\v{c}}nik}, \citenamefont {Wallraff} \emph {et~al.}}]{heinsoo2018rapid}%
  \BibitemOpen
  \bibfield  {author} {\bibinfo {author} {\bibfnamefont {J.}~\bibnamefont {Heinsoo}}, \bibinfo {author} {\bibfnamefont {C.~K.}\ \bibnamefont {Andersen}}, \bibinfo {author} {\bibfnamefont {A.}~\bibnamefont {Remm}}, \bibinfo {author} {\bibfnamefont {S.}~\bibnamefont {Krinner}}, \bibinfo {author} {\bibfnamefont {T.}~\bibnamefont {Walter}}, \bibinfo {author} {\bibfnamefont {Y.}~\bibnamefont {Salath{\'e}}}, \bibinfo {author} {\bibfnamefont {S.}~\bibnamefont {Gasparinetti}}, \bibinfo {author} {\bibfnamefont {J.-C.}\ \bibnamefont {Besse}}, \bibinfo {author} {\bibfnamefont {A.}~\bibnamefont {Poto{\v{c}}nik}}, \bibinfo {author} {\bibfnamefont {A.}~\bibnamefont {Wallraff}}, \emph {et~al.},\ }\bibfield  {title} {\bibinfo {title} {Rapid high-fidelity multiplexed readout of superconducting qubits},\ }\href {https://link.aps.org/doi/10.1103/PhysRevApplied.10.034040} {\bibfield  {journal} {\bibinfo  {journal} {Phys. Rev. Appl.}\ }\textbf {\bibinfo {volume} {10}},\ \bibinfo {pages} {034040} (\bibinfo {year}
  {2018})}\BibitemShut {NoStop}%
\bibitem [{\citenamefont {Cleland}\ \emph {et~al.}(2019)\citenamefont {Cleland}, \citenamefont {Pechal}, \citenamefont {Stas}, \citenamefont {Sarabalis}, \citenamefont {Wollack},\ and\ \citenamefont {Safavi-Naeini}}]{cleland2019mechanical}%
  \BibitemOpen
  \bibfield  {author} {\bibinfo {author} {\bibfnamefont {A.~Y.}\ \bibnamefont {Cleland}}, \bibinfo {author} {\bibfnamefont {M.}~\bibnamefont {Pechal}}, \bibinfo {author} {\bibfnamefont {P.-J.~C.}\ \bibnamefont {Stas}}, \bibinfo {author} {\bibfnamefont {C.~J.}\ \bibnamefont {Sarabalis}}, \bibinfo {author} {\bibfnamefont {E.~A.}\ \bibnamefont {Wollack}},\ and\ \bibinfo {author} {\bibfnamefont {A.~H.}\ \bibnamefont {Safavi-Naeini}},\ }\bibfield  {title} {\bibinfo {title} {Mechanical purcell filters for microwave quantum machines},\ }\href {https://doi.org/10.1063/1.5111151} {\bibfield  {journal} {\bibinfo  {journal} {Appl. Phys. Lett.}\ }\textbf {\bibinfo {volume} {115}} (\bibinfo {year} {2019})}\BibitemShut {NoStop}%
\bibitem [{\citenamefont {C{\'o}rcoles}\ \emph {et~al.}(2011)\citenamefont {C{\'o}rcoles}, \citenamefont {Chow}, \citenamefont {Gambetta}, \citenamefont {Rigetti}, \citenamefont {Rozen}, \citenamefont {Keefe}, \citenamefont {Beth~Rothwell}, \citenamefont {Ketchen},\ and\ \citenamefont {Steffen}}]{corcoles2011protecting}%
  \BibitemOpen
  \bibfield  {author} {\bibinfo {author} {\bibfnamefont {A.~D.}\ \bibnamefont {C{\'o}rcoles}}, \bibinfo {author} {\bibfnamefont {J.~M.}\ \bibnamefont {Chow}}, \bibinfo {author} {\bibfnamefont {J.~M.}\ \bibnamefont {Gambetta}}, \bibinfo {author} {\bibfnamefont {C.}~\bibnamefont {Rigetti}}, \bibinfo {author} {\bibfnamefont {J.~R.}\ \bibnamefont {Rozen}}, \bibinfo {author} {\bibfnamefont {G.~A.}\ \bibnamefont {Keefe}}, \bibinfo {author} {\bibfnamefont {M.}~\bibnamefont {Beth~Rothwell}}, \bibinfo {author} {\bibfnamefont {M.~B.}\ \bibnamefont {Ketchen}},\ and\ \bibinfo {author} {\bibfnamefont {M.}~\bibnamefont {Steffen}},\ }\bibfield  {title} {\bibinfo {title} {Protecting superconducting qubits from radiation},\ }\href {https://doi.org/10.1063/1.3658630} {\bibfield  {journal} {\bibinfo  {journal} {Appl. Phys. Lett.}\ }\textbf {\bibinfo {volume} {99}} (\bibinfo {year} {2011})}\BibitemShut {NoStop}%
\bibitem [{\citenamefont {Harrington}\ \emph {et~al.}(2022)\citenamefont {Harrington}, \citenamefont {Mueller},\ and\ \citenamefont {Murch}}]{harrington2022engineered}%
  \BibitemOpen
  \bibfield  {author} {\bibinfo {author} {\bibfnamefont {P.~M.}\ \bibnamefont {Harrington}}, \bibinfo {author} {\bibfnamefont {E.~J.}\ \bibnamefont {Mueller}},\ and\ \bibinfo {author} {\bibfnamefont {K.~W.}\ \bibnamefont {Murch}},\ }\bibfield  {title} {\bibinfo {title} {Engineered dissipation for quantum information science},\ }\href {https://doi.org/https://doi.org/10.1038/s42254-022-00494-8} {\bibfield  {journal} {\bibinfo  {journal} {Nat. Rev. Phys.}\ }\textbf {\bibinfo {volume} {4}},\ \bibinfo {pages} {660} (\bibinfo {year} {2022})}\BibitemShut {NoStop}%
\bibitem [{\citenamefont {Shankar}\ \emph {et~al.}(2013)\citenamefont {Shankar}, \citenamefont {Hatridge}, \citenamefont {Leghtas}, \citenamefont {Sliwa}, \citenamefont {Narla}, \citenamefont {Vool}, \citenamefont {Girvin}, \citenamefont {Frunzio}, \citenamefont {Mirrahimi},\ and\ \citenamefont {Devoret}}]{shankar2013autonomously}%
  \BibitemOpen
  \bibfield  {author} {\bibinfo {author} {\bibfnamefont {S.}~\bibnamefont {Shankar}}, \bibinfo {author} {\bibfnamefont {M.}~\bibnamefont {Hatridge}}, \bibinfo {author} {\bibfnamefont {Z.}~\bibnamefont {Leghtas}}, \bibinfo {author} {\bibfnamefont {K.}~\bibnamefont {Sliwa}}, \bibinfo {author} {\bibfnamefont {A.}~\bibnamefont {Narla}}, \bibinfo {author} {\bibfnamefont {U.}~\bibnamefont {Vool}}, \bibinfo {author} {\bibfnamefont {S.~M.}\ \bibnamefont {Girvin}}, \bibinfo {author} {\bibfnamefont {L.}~\bibnamefont {Frunzio}}, \bibinfo {author} {\bibfnamefont {M.}~\bibnamefont {Mirrahimi}},\ and\ \bibinfo {author} {\bibfnamefont {M.~H.}\ \bibnamefont {Devoret}},\ }\bibfield  {title} {\bibinfo {title} {Autonomously stabilized entanglement between two superconducting quantum bits},\ }\href {https://doi.org/https://doi.org/10.1038/nature12802} {\bibfield  {journal} {\bibinfo  {journal} {Nature}\ }\textbf {\bibinfo {volume} {504}},\ \bibinfo {pages} {419} (\bibinfo {year} {2013})}\BibitemShut {NoStop}%
\bibitem [{\citenamefont {Kerckhoff}\ \emph {et~al.}(2010)\citenamefont {Kerckhoff}, \citenamefont {Nurdin}, \citenamefont {Pavlichin},\ and\ \citenamefont {Mabuchi}}]{kerckhoff2010designing}%
  \BibitemOpen
  \bibfield  {author} {\bibinfo {author} {\bibfnamefont {J.}~\bibnamefont {Kerckhoff}}, \bibinfo {author} {\bibfnamefont {H.~I.}\ \bibnamefont {Nurdin}}, \bibinfo {author} {\bibfnamefont {D.~S.}\ \bibnamefont {Pavlichin}},\ and\ \bibinfo {author} {\bibfnamefont {H.}~\bibnamefont {Mabuchi}},\ }\bibfield  {title} {\bibinfo {title} {Designing quantum memories with embedded control: Photonic circuits for autonomous quantum error correction},\ }\href {https://doi.org/10.1103/PhysRevLett.105.040502} {\bibfield  {journal} {\bibinfo  {journal} {Phys. Rev. Lett.}\ }\textbf {\bibinfo {volume} {105}},\ \bibinfo {pages} {040502} (\bibinfo {year} {2010})}\BibitemShut {NoStop}%
\bibitem [{\citenamefont {Kapit}(2016)}]{kapit2016hardware}%
  \BibitemOpen
  \bibfield  {author} {\bibinfo {author} {\bibfnamefont {E.}~\bibnamefont {Kapit}},\ }\bibfield  {title} {\bibinfo {title} {Hardware-efficient and fully autonomous quantum error correction in superconducting circuits},\ }\href {https://doi.org/10.1103/PhysRevLett.116.150501} {\bibfield  {journal} {\bibinfo  {journal} {Phys. Rev. Lett.}\ }\textbf {\bibinfo {volume} {116}},\ \bibinfo {pages} {150501} (\bibinfo {year} {2016})}\BibitemShut {NoStop}%
\bibitem [{\citenamefont {Reiter}\ \emph {et~al.}(2017)\citenamefont {Reiter}, \citenamefont {S{\o}rensen}, \citenamefont {Zoller},\ and\ \citenamefont {Muschik}}]{reiter2017dissipative}%
  \BibitemOpen
  \bibfield  {author} {\bibinfo {author} {\bibfnamefont {F.}~\bibnamefont {Reiter}}, \bibinfo {author} {\bibfnamefont {A.~S.}\ \bibnamefont {S{\o}rensen}}, \bibinfo {author} {\bibfnamefont {P.}~\bibnamefont {Zoller}},\ and\ \bibinfo {author} {\bibfnamefont {C.}~\bibnamefont {Muschik}},\ }\bibfield  {title} {\bibinfo {title} {Dissipative quantum error correction and application to quantum sensing with trapped ions},\ }\href {https://doi.org/https://doi.org/10.1038/s41467-017-01895-5} {\bibfield  {journal} {\bibinfo  {journal} {Nat. Commun.}\ }\textbf {\bibinfo {volume} {8}},\ \bibinfo {pages} {1822} (\bibinfo {year} {2017})}\BibitemShut {NoStop}%
\bibitem [{\citenamefont {Gertler}\ \emph {et~al.}(2021)\citenamefont {Gertler}, \citenamefont {Baker}, \citenamefont {Li}, \citenamefont {Shirol}, \citenamefont {Koch},\ and\ \citenamefont {Wang}}]{gertler2021protecting}%
  \BibitemOpen
  \bibfield  {author} {\bibinfo {author} {\bibfnamefont {J.~M.}\ \bibnamefont {Gertler}}, \bibinfo {author} {\bibfnamefont {B.}~\bibnamefont {Baker}}, \bibinfo {author} {\bibfnamefont {J.}~\bibnamefont {Li}}, \bibinfo {author} {\bibfnamefont {S.}~\bibnamefont {Shirol}}, \bibinfo {author} {\bibfnamefont {J.}~\bibnamefont {Koch}},\ and\ \bibinfo {author} {\bibfnamefont {C.}~\bibnamefont {Wang}},\ }\bibfield  {title} {\bibinfo {title} {Protecting a bosonic qubit with autonomous quantum error correction},\ }\href {https://doi.org/https://doi.org/10.1038/s41586-021-03257-0} {\bibfield  {journal} {\bibinfo  {journal} {Nature}\ }\textbf {\bibinfo {volume} {590}},\ \bibinfo {pages} {243} (\bibinfo {year} {2021})}\BibitemShut {NoStop}%
\bibitem [{\citenamefont {Marquet}\ \emph {et~al.}(2024)\citenamefont {Marquet}, \citenamefont {Essig}, \citenamefont {Cohen}, \citenamefont {Cottet}, \citenamefont {Murani}, \citenamefont {Albertinale}, \citenamefont {Dupouy}, \citenamefont {Bienfait}, \citenamefont {Peronnin}, \citenamefont {Jezouin}, \citenamefont {Lescanne},\ and\ \citenamefont {Huard}}]{Marquet2024Autoparametric}%
  \BibitemOpen
  \bibfield  {author} {\bibinfo {author} {\bibfnamefont {A.}~\bibnamefont {Marquet}}, \bibinfo {author} {\bibfnamefont {A.}~\bibnamefont {Essig}}, \bibinfo {author} {\bibfnamefont {J.}~\bibnamefont {Cohen}}, \bibinfo {author} {\bibfnamefont {N.}~\bibnamefont {Cottet}}, \bibinfo {author} {\bibfnamefont {A.}~\bibnamefont {Murani}}, \bibinfo {author} {\bibfnamefont {E.}~\bibnamefont {Albertinale}}, \bibinfo {author} {\bibfnamefont {S.}~\bibnamefont {Dupouy}}, \bibinfo {author} {\bibfnamefont {A.}~\bibnamefont {Bienfait}}, \bibinfo {author} {\bibfnamefont {T.}~\bibnamefont {Peronnin}}, \bibinfo {author} {\bibfnamefont {S.}~\bibnamefont {Jezouin}}, \bibinfo {author} {\bibfnamefont {R.}~\bibnamefont {Lescanne}},\ and\ \bibinfo {author} {\bibfnamefont {B.}~\bibnamefont {Huard}},\ }\bibfield  {title} {\bibinfo {title} {Autoparametric resonance extending the bit-flip time of a cat qubit up to 0.3 s},\ }\href {https://link.aps.org/doi/10.1103/PhysRevX.14.021019} {\bibfield  {journal} {\bibinfo  {journal} {Phys. Rev.
  X}\ }\textbf {\bibinfo {volume} {14}},\ \bibinfo {pages} {021019} (\bibinfo {year} {2024})}\BibitemShut {NoStop}%
\bibitem [{\citenamefont {Harris}\ \emph {et~al.}(2017)\citenamefont {Harris}, \citenamefont {Steinbrecher}, \citenamefont {Prabhu}, \citenamefont {Lahini}, \citenamefont {Mower}, \citenamefont {Bunandar}, \citenamefont {Chen}, \citenamefont {Wong}, \citenamefont {Baehr-Jones}, \citenamefont {Hochberg} \emph {et~al.}}]{harris2017quantum}%
  \BibitemOpen
  \bibfield  {author} {\bibinfo {author} {\bibfnamefont {N.~C.}\ \bibnamefont {Harris}}, \bibinfo {author} {\bibfnamefont {G.~R.}\ \bibnamefont {Steinbrecher}}, \bibinfo {author} {\bibfnamefont {M.}~\bibnamefont {Prabhu}}, \bibinfo {author} {\bibfnamefont {Y.}~\bibnamefont {Lahini}}, \bibinfo {author} {\bibfnamefont {J.}~\bibnamefont {Mower}}, \bibinfo {author} {\bibfnamefont {D.}~\bibnamefont {Bunandar}}, \bibinfo {author} {\bibfnamefont {C.}~\bibnamefont {Chen}}, \bibinfo {author} {\bibfnamefont {F.~N.}\ \bibnamefont {Wong}}, \bibinfo {author} {\bibfnamefont {T.}~\bibnamefont {Baehr-Jones}}, \bibinfo {author} {\bibfnamefont {M.}~\bibnamefont {Hochberg}}, \emph {et~al.},\ }\bibfield  {title} {\bibinfo {title} {Quantum transport simulations in a programmable nanophotonic processor},\ }\href {https://doi.org/https://doi.org/10.1038/nphoton.2017.95} {\bibfield  {journal} {\bibinfo  {journal} {Nature Photon.}\ }\textbf {\bibinfo {volume} {11}},\ \bibinfo {pages} {447} (\bibinfo {year} {2017})}\BibitemShut
  {NoStop}%
\end{thebibliography}%

\end{document}